\documentclass[12pt]{amsart}

\newcommand{\commentout}[1]{}

\reversemarginpar
\usepackage{amsmath}
\usepackage{amssymb}

\newcommand{\nwc}{\newcommand}

\newcommand{\lt}{\left}
\nwc{\partz}{\frac{\partial }{\partial z}}
\newcommand{\rt}{\right}
\nwc{\ytil}{\tilde{\by}}
\nwc{\al}{\alpha}
\newcommand{\lan}{\left\langle}
\newcommand{\ran}{\right\rangle}

\newcommand{\bx}{\mathbf x}

\newcommand{\mbR}{\mathbf{R}}
\nwc{\fR}{\frak{R}}
\newcommand{\bp}{\mathbf p}

\newcommand{\bq}{\mathbf q}
\newcommand{\by}{\mathbf y}

\newcommand{\cpv}{\!\!\!\!\!\! - \  }
\nwc{\nwt}{\newtheorem}

\nwt{proposition}{Proposition}
\nwt{lemma}{Lemma}
\nwt{theorem}{Theorem}
\nwt{remark}{Remark}
\nwt{assumption}{Assumption}
\nwt{definition}{Definition} %def is already defined

\nwc{\bal}{\begin{align}}
\nwc{\be}{\begin{equation}}
\nwc{\ben}{\begin{equation*}}
\nwc{\bea}{\begin{eqnarray}}
\nwc{\beq}{\begin{eqnarray}}
\nwc{\bean}{\begin{eqnarray*}}
\nwc{\beqn}{\begin{eqnarray*}}
\nwc{\beqast}{\begin{eqnarray*}}

\nwc{\eal}{\end{align}}
\nwc{\ee}{\end{equation}}
\nwc{\een}{\end{equation*}}
\nwc{\eea}{\end{eqnarray}}
\nwc{\eeq}{\end{eqnarray}}
\nwc{\eean}{\end{eqnarray*}}
\nwc{\eeqn}{\end{eqnarray*}}
\nwc{\eeqast}{\end{eqnarray*}}

\nwc{\invf}{\cF^{-1}_2}
\nwc{\ep}{\ell}
\nwc{\tep}{\tilde{\varepsilon}}
\nwc{\epsq}{{\varepsilon^2}}
\nwc{\epsqa}{{\varepsilon^{2\alpha}}}
\nwc{\eps}{\varepsilon}
\nwc{\ept}{\epsilon}
\nwc{\vrho}{\varrho}
\nwc{\orho}{\bar\varrho}
\nwc{\ou}{\bar u}
\nwc{\vpsi}{\varpsi}
\nwc{\lamb}{\ell}

\nwc{\nn}{\nonumber}
\nwc{\bm}{\boldmath}
\nwc{\mf}{\mathbf}
\nwc{\mb}{\mathbf}
\nwc{\ml}{\mathcal}

\nwc{\IA}{\mathbb{A}} %algebraic
\nwc{\IB}{\mathbb{B}}
\nwc{\IC}{\mathbb{C}} %complex
\nwc{\ID}{\mathbb{D}} %Dedekind
\nwc{\IM}{\mathbb{M}} %Dedekind
\nwc{\IP}{\mathbb{P}} %Dedekind
\nwc{\II}{\mathbb{I}} %Dedekind
\nwc{\IE}{\mathbb{E}} %Euklides
\nwc{\IF}{\mathbb{F}} %finite field
\nwc{\IG}{\mathbb{G}} %Gauss
\nwc{\IN}{\mathbb{N}} %natural
\nwc{\IQ}{\mathbb{Q}} %rational
\nwc{\IR}{\mathbb{R}} %real
\nwc{\IT}{\mathbb{T}} %torus
\nwc{\IZ}{\mathbb{Z}} %integers

\nwc{\epal}{\ep^{-2\alpha}}
\nwc{\cE}{{\ml E}}
\nwc{\cP}{{\ml P}}
\nwc{\cQ}{{\ml Q}}
\nwc{\cL}{{\ml L}}
\nwc{\cR}{{\ml R}}
\nwc{\cV}{{\ml V}}
\nwc{\cT}{{\ml T}}
\nwc{\crV}{{\ml L}_{(\delta,\nu)}}
\nwc{\cC}{{\ml C}}
\nwc{\cA}{{\ml A}}
\nwc{\cK}{{\ml S}}
\nwc{\cB}{{\ml B}}
\nwc{\cD}{{\ml D}}
\nwc{\cF}{{\ml F}}
\nwc{\cS}{{\ml S}}
\nwc{\cM}{{\ml M}}
\nwc{\cG}{{\ml G}}
\nwc{\cH}{{\ml H}}
\nwc{\bk}{{\mb k}}
\nwc{\cbz}{\overline{\cB}_z}
\nwc{\bE}{{\mb E}}
\nwc{\bH}{{\mb H}}
\nwc{\bO}{{\mb O}}
\nwc{\bU}{{\mb U}}
\nwc{\bW}{{\mb W}}
\nwc{\bK}{{\mb K}}
\nwc{\bI}{{\mb I}}
\nwc{\bP}{{\mb P}}
\nwc{\bV}{{\mb V}}
\nwc{\bb}{{\mb e}}
\nwc{\bba}{{{\mb e}^{\sigma,\alpha}}}
\nwc{\bbz}{{{\mb e}^{\sigma,\zeta}}}
\nwc{\bbe}{{{\mb e}^{\sigma,\eta}}}
\nwc{\bB}{{\mb B}}
\nwc{\bEaz}{{{\mb E}^{\sigma,\alpha\zeta}}}
\nwc{\bF}{{\mb F}}
\nwc{\bG}{{\mb G}}
\nwc{\bT}{{\mb T}}
\nwc{\Waz}{{\bar W}^\sigma_{\alpha\zeta}}
\nwc{\Caz}{{C}^\sigma_{\alpha\zeta}}
\nwc{\bd}{{\mb d}}
\nwc{\bda}{{{\mb d}^{\sigma,\alpha}}}
\nwc{\bdz}{{{\mb d}^{\sigma,\zeta}}}
\nwc{\bde}{{{\mb d}^{\sigma,\eta}}}
\nwc{\bD}{{\mb D}}
\nwc{\bDaz}{{{\mb D}^{\sigma,\alpha\zeta}}}
\nwc{\Om}{{\Omega}}
\nwc{\bSig}{{\mb \Sigma}}
\nwc{\fM}{\frak{M}}
\nwc{\bJ}{{\mb J}}
\nwc{\fH}{\frak{H}}
\nwc{\fQ}{\frak{P}}
\nwc{\ga}{\gamma}
\nwc{\bX}{{\mb X}}
\nwc{\fX}{\frak{X}}
\nwc{\bY}{{\mb Y}}
\nwc{\fY}{\frak{Y}}
\nwc{\fC}{\frak{C}}
\nwc{\bn}{\mb n}
\nwc{\bu}{\mb u}
\nwc{\bg}{{\mb g}}
\nwc{\bL}{{\mb L}}
\nwc{\ppj}{\partial_{p_{j}}}
\nwc{\pxj}{\partial_{x_{j}}}
\nwc{\pxjtil}{\partial_{\tilde x_{j}}}
\nwc{\ppl}{\partial_{p_{l}}}
\nwc{\pxl}{\partial_{x_{l}}}
\nwc{\bC}{\mb C}
\nwc{\dagg}{\dagger}
\nwc{\partpl}{\partial_{p_l}}
\nwc{\partxl}{\partial_{x_l}}

\nwc{\pft}{\cF^{-1}_\bp}

\newcommand{\fW}{\mathfrak{W}}
\newcommand{\om}{\omega}
\nwc{\fS}{\frak{S}}
\nwc{\fF}{\frak{F}}
\nwc{\bQ}{{\mb P}}
\begin{document}

%%%%%%%%%%%%%%%%%% title page information %%%%%%%%%%%%%%%%%%
\title{Mutual Coherence of Polarized Light in Disordered Media: Two-Frequency Method Extended}

\author{Albert C. Fannjiang}
 \thanks{Department of Mathematics,
University of California,
Davis, CA 95616-8633. Email: fannjiang@math.ucdavis.edu
The research is supported in part by the Defense Advanced Research Projects Agency (DARPA) grant 
 N00014-02-1-0603
}

%%%%%%%%%%%%%%%%%%%%%%
\commentout{
\begin{frontmatter}\title{Two-Point Correlations of Polarized Light in Disordered Media: Two-Frequency Method Extended}
\author{Albert C. Fannjiang}
 \ead{
  fannjiang@math.ucdavis.edu}
   \ead[url]{http://www.math.ucdavis.edu/\~\,fannjian}
 \thanks{
The research is supported in part by the Defense Advanced Research Projects Agency (DARPA) grant 
 N00014-02-1-0603
}
 \address[al]{
Department of Mathematics,
University of California, Davis 95616-8633}
}

\begin{abstract}
The paper addresses  the two-point
correlations of electromagnetic waves in general
random, bi-anisotropic media  whose constitutive 
tensors  are complex Hermitian, positive- or negative-definite matrices. A simplified version of the two-frequency
Wigner distribution  (2f-WD) for polarized waves is introduced
and the closed form Wigner-Moyal equation is
derived from the Maxwell equations. In the weak-disorder regime
with an {\em arbitrarily varying } background
the two-frequency radiative transfer (2f-RT) equations
for the associated $2\times 2$ coherence matrices
are derived  from the Wigner-Moyal equation by  using the multiple scale expansion.  In birefringent media,
the coherence matrix becomes a scalar and 
the 2f-RT equations take the scalar form due to the absence of
depolarization. 
A paraxial approximation is developed for
 spatialy anisotropic media. Examples
 of isotropic, chiral, uniaxial and gyrotropic media are
 discussed. 
 
 \bigskip
 
\noindent  PACS numbers: 42.25.Dd, 41.20.Jb
\end{abstract}

%\ocis{030.5620, 290.4210} 

%\end{frontmatter}
\maketitle

\section{Introduction}

Consider the electromagnetic wave propagation in a random
dielectric. Let $\bu(\bx,t)=(\bD(\bx,t),\mb{B}(\bx,t))^\dagger$ be the displacement-magnetic-induction vector field. Then the
mutual coherence function  is given by \cite{BW}
\beq
\lan \bu(\bx_{1},t_{1})\bu^{\dagger}(\bx_{2},t_{2})\ran
=\int e^{i(\om_{2}-\om_{1})t} e^{-i\tau(\om_{1}+\om_{2})/2} \lan\bU(\bx_{1}, \om_{1})\bU^{\dagger}(\bx_{2}, \om_{2})\ran d\om_{1}
d\om_{2}\label{2st}
\eeq
where $\lan\cdot\ran$ is ensemble averaging, $t=(t_{1}+t_{2})/2, \tau=t_{1}-t_{2}$ and
$\bU(\bx, \om)$ is the frequency component of $\bu$ at
frequency $\om$. Throughout, all vectors are by default
column vectors and $\dagger$ denotes Hermitian conjugation.

Radiative transfer theory \cite{BVKT, Cha, Fan, Kok,  LW, MTL, PB, RPK} has been traditionally carried
out in space-time  with one time variable ($t_{1}=t_{2}$).
The main goal of this paper is to derive equations for 
the quantity directly related to $\lan\bU(\bx_{1}, \om_{1})\bU^{\dagger}(\bx_{2}, \om_{2})\ran$.
In particular, we obtain the two-frequency
radiative transfer (2f-RT) equations for polarized light
in the weak-disorder regime with an arbitrary bianisotropic background. The 2f-RT equations then determine, via (\ref{2st}), the two-space-time correlations of the electromagnetic wave in random media. 

Our approach is set up for the most general 
 linear local, lossless electromagnetic
materials, in which each of the field vectors $\bE$ and $\bH$ is coupled
tensorially to both  $\bD$ and $\mathbf{B}$. Such materials have
been the subject of considerable recent interest.
One reason is that they can be
created as metamaterials, i.e. composites of more conventional materials in which
$\bE$ is coupled to $\bD$ alone and $\bH$ is coupled to $\mb{B}$ alone. 

The two-frequency approach has been previously pursued  in terms of wavelength-rescaled two-frequency Wigner
distribution (2f-WD) in the case of a {\em uniform}  background
\cite{2frt-maxwell}. In the present work, we introduce
an alternative version of 2f-WD and derive
 the corresponding 2f-RT equations for 
 the associated $2\times 2$ coherence matrices in 
the case with an {\em arbitrary} background. We give several
examples for which the scattering kernels can be computed
explicitly. We show that birefringence naturally leads to
decoupling of the polarization modes and the absence
of depolarization in such media. As a result, the 2f-RT equations
simplify to scalar equations. 

In Section \ref{sec2} we formulate the problem in terms of 
the straightforwardly defined 2f-WD and derive
the two-frequency Wigner-Moyal equation in Appendix A. In Section \ref{sec:go}
we analyze the problem for  high-frequency waves in an arbitrary background bianisotropic medium
in the absence of random fluctuations. This is the geometrical optics regime. In Section \ref{sec3}, we consider the weak-disorder regime where, in addition to the arbitrary background,
 small random fluctuations are present on the scale
 of the wavelength. In Section \ref{sec4} and \ref{sec5} 
 we employ the multiscale expansion technique to
 derive the radiative transfer equation from
 the two-frequency Wigner-Moyal equation
 for the weak-disorder regime. We also derive
 a paraxial approximation for the polarized light
 in a spatially anisotropic medium. 
  In Section \ref{sec:med}
 we give several examples of isotropic, chiral, uniaxial and gyrotropic media for which the scattering
 kernels can be explicitly calculated.  
 We conclude in Section~\ref{con} with
 a discussion of the final expression of
 the mutual coherence in terms of the solution
 of the 2f-RT equations.

\section{Maxwell equations and Wigner-Moyal equations}
\label{sec2}
In this paper, we consider the electromagnetic wave propagation
%(in the {\em m.k.s.a} system of units) 
in a heterogeneous, lossless, bi-anisotropic dielectric medium.
We assume that the scattering medium is free of charges and currents and start with 
the source-free Maxwell equations in the frequency  $\om$ domain
\beq
-i\omega\lt[\begin{matrix}
 \bD \\
{\mb B}
\end{matrix}\rt]+\lt[
\begin{matrix}
0 &-\nabla\times\\
\nabla \times &0
\end{matrix}
\rt]\bK^{-1}\lt[\begin{matrix}
\bD\\
{\mb B}
\end{matrix}
\rt]=0\label{max}
\eeq
where $\bK$ is, by the assumption of losslessness, a Hermitian matrix \cite{LST}
\beq
\label{K}
\bK=\lt[\begin{matrix}
\bK^\epsilon&\bK^\chi\\
\bK^{\chi\dagg}&\bK^\mu
\end{matrix}
\rt]
\eeq
%where $\bE$ and $\bH$ are the electric and magnetic fields
%respectively, 
with 
  the permittivity   and 
permeability tensors $\bK^\epsilon, \bK^\mu$, and  the
magneto-electric tensor $\bK^\chi$ \cite{Ode}. 
The Hermitian matrix  $\bK$ is assumed to be invertible.
%Although we do not assume the positive-definiteness of $\bK$
%The material coefficients in the so-called Tellegen constitutive relations must
%satisfy the constraint
%\[
%\hbox{Tr}\Big[\big(\bK^\mu\big)^{-1}(\bK^\xi+\bK^\zeta)\Big]=0.
%\]
The present formulation encompasses the acoustic, electromagnetic and elastic waves so that the 2f-RT
theory developed here can be extended to these waves 
without major changes.  
We choose
$\bD, {\mb B}$ as the primary fields because they
are transverse (divergence-free).

In an isotropic dielectric, $\bK^\epsilon=\epsilon\bI, \bK^\mu=\mu\bI, \bK^\chi=0$. In a biisotropic dielectric,
$\bK^\chi$ as well as $\bK^{\epsilon}, \bK^{\mu}$ are nonzero scalars. A reciprocal chiral medium is biisotropic with purely
imaginary $\bK^{\chi}=i\chi $. The appearance of nonzero $\bK^\chi$ arises from
the so called magnetoelectric effect \cite{LLP}.
Crystals are often naturally anisotropic, and in some media (such as liquid crystals) it is possible to induce anisotropy by applying e.g. an external electric field. In crystal optics,  $\bK^\epsilon,\bK^\mu$ are  real, symmetric matrices 
and $\bK^\chi=0$ \cite{BW}. 
In response to a magnetic field, some materials can have a dielectric tensor that is complex-Hermitian; this is the gyrotropic  effect. A magnetoelectric, bi-anisotropic medium has
a constitutive relation  (\ref{K}) with complex Hermitian $\bK^{\epsilon}, \bK^{\mu}$ and
a complex matrix $\bK^{\chi}$ satisfying
the Post constraint. It has been shown that a moving medium, even isotropic, must be treated as bi-anisotropic
\cite{CK,LLP}.

Writing the total field $\bU=(\bD, {\mb B})$ we introduce the two-frequency
matrix-valued Wigner distribution
\beq
\label{1.22}
\bW(\bx,\bp;\omega_1,\omega_2)
=\frac{1}{(2\pi)^3}
\int e^{-i\bp^\dagg\by} \bU_1\big({\bx}+\frac{\ell\by}{2}\big)
\bU_2^\dagg\big({\bx}-\frac{\ell\by}{2}\big)d\by
\eeq
where $\bU_1$ and $\bU_2$ are the total fields
at frequencies $\omega_1/\ell$ and $ \omega_2/\ell$ respectively.
The parameter $\ell$ is roughly  the ratio of the wavelength to
the distance of propagation. In the present setting, $\ell\ll 1$.
Correspondingly, we will  replace $\om$ in (\ref{max}) by
$\om/\ell$. The 2f-WD is clearly equivalent to
the 2-point function $\bU_{1}\bU_{2}^{\dagger}$ via
the inverse Fourier transform.

Notice the symmetry of the Wigner distribution matrix
\beq
\label{4}
\bW^\dagg(\bx,\bp;\omega_1,\omega_2)=\bW(\bx,\bp; \omega_2, \omega_1).
\eeq
 In other words, the right hand side of
(\ref{1.22}) is invariant under the simultaneous  transformations
of 
Hermitian conjugation  $\dagger$ and frequency exchange $\omega_1\leftrightarrow \omega_2$.

In what follows we shall omit writing the arguments of
any fields unless necessary. 

We put the equation (\ref{max})
in the form of general symmetric hyperbolic system \cite{RPK}
\beq
\label{hyper}
-i\frac{\om}{\ell} \bU+\mbR_{j}\pxj\lt( \bK^{-1}\bU\rt)=0
\eeq
where  $\mbR_{j}$ are the symmetric matrix  given by
\beqn
\mbR_j=\lt[\begin{matrix}
0&\bT_j\\
-\bT_j&0
\end{matrix}
\rt]
\eeqn
with
\beqn
\bT_1=\lt[\begin{matrix}
0&0&0\\
0&0&-1\\
0&1&0
\end{matrix}
\rt],\quad
\bT_2=\lt[\begin{matrix}
0&0&1\\
0&0&0\\
-1&0&0
\end{matrix}
\rt],\quad\bT_3=\lt[\begin{matrix}
0&-1&0\\
1&0&0\\
0&0&0
\end{matrix}
\rt]. 
\eeqn
The matrices $iT_j, j=1,2,3$ are related to
the photon spin matrices \cite{BB}. 
For ease of notation, we set $\bL=\bK^{-1}$. 

Let $\om'=(\om_{1}-\om_{2})/\ell$ and
$\bar\om=(\om_{1}+\om_{2})/2$. The 2f-WD satisfies the Wigner-Moyal equations
\beq
\label{wig-mo}
i\om'\bW
&=&\frac{i}{\ell}  p_j\mbR_j\int e^{i\bq^{\dagger}\bx}\widehat\bL(\bq) 
\bW(\bp-\frac{\ell\bq}{2})d\bq
-\frac{i}{\ell}\int \bW(\bp+\frac{\ell\bq}{2})
\widehat\bL(\bq)e^{i\bq^{\dagger}\bx}d\bq  p_j\mbR_j\\
&&+\frac{1}{2} \mbR_j\pxj
\int e^{i\bq^{\dagger}\bx}\widehat\bL(\bq)\bW(\bp-\frac{\ell\bq}{2})
d\bq+\frac{1}{2}\pxj\int
\bW(\bp+\frac{\ell\bq}{2})\widehat\bL(\bq)e^{i\bq^{\dagger}\bx}d\bq
\mbR_{j}\nn\\
i\frac{2\bar\om}{\ell}\bW
&=&\frac{i}{\ell}  p_j\mbR_j\int e^{i\bq^{\dagger}\bx}\widehat\bL(\bq) 
\bW(\bp-\frac{\ell\bq}{2})d\bq
+\frac{i}{\ell}\int \bW(\bp+\frac{\ell\bq}{2})
\widehat\bL(\bq)e^{i\bq^{\dagger}\bx}d\bq  p_j\mbR_j \label{wig-mo2}\\
&&+\frac{1}{2} \mbR_j\pxj
\int e^{i\bq^{\dagger}\bx}\widehat\bL(\bq)\bW(\bp-\frac{\ell\bq}{2})
d\bq-\frac{1}{2}\pxj\int
\bW(\bp+\frac{\ell\bq}{2})\widehat\bL(\bq)e^{i\bq^{\dagger}\bx}d\bq
\mbR_{j}\nn
\eeq
where $\widehat\bL$ is the Fourier transform (spectral density) of
$\bL$ 
\[
\bL(\bx)=\int e^{i\bx^\dagger\bq}\widehat \bL(\bq)  d\bq.
\]
The derivation is given in Appendix A.
 For a Hermitian $\bL$ we have
\[
\hat\bL(\bp)=\hat\bL^\dagger(-\bp),\quad\forall \bp.
\]
Clearly eq. (\ref{wig-mo}) is related to
the time derivative of the mutual coherence (\ref{2st})
with respect to the central time $t$
while eq. (\ref{wig-mo}) is related to
the time derivative with respect to the differential time $\tau$.
Our  will focus first on eq. (\ref{wig-mo})
and comment on the constraint posed by (\ref{wig-mo2})
in the Conclusion. 
The full analysis of eq.  (\ref{wig-mo2}) requires  substantially
different treatment and will be presented  elsewhere.  
However, we will discuss the constraint imposed by
the leading order terms of eq. (\ref{wig-mo2}) in
the Conclusion and its implication on the two-spacetime 
correlation.

\section{Geometrical optics}
\label{sec:go}
In this regime, we let $\ell\ll 1$ implying  a small ratio between the wavelength
and the scale of background heterogeneity which is
comparable to the distance of propagation. 

Let us first
simplify eq. (\ref{wig-mo}) by expanding the expression in the power of $\ell$ and neglecting $O(\ell)$-terms. 
The first two terms on the right hand side of eq. (\ref{wig-mo}) reduce to $\ell^{-1}\cP_{0}-\cP_{1}$ where 
\beqn
\cP_{0}(\bp)\bW&=&{i}  p_j\mbR_j\bL\bW-{i}\bW\bL  p_j\mbR_j\\
\cP_{1}(\bp)\bW&=&
\frac{1}{2}  p_j\mbR_j \pxl\bL \partial_{p_{l}}\bW
+\frac{1}{2}\partial_{p_{l}}\bW\pxl\bL
  p_j\mbR_j
\eeqn
while the last two terms on the right hand side become
\beq
\cP_{2}\bW=\frac{1}{2} \mbR_j\pxj
\lt[\bL\bW\rt]+\frac{1}{2}\pxj
\lt[\bW \bL\rt]
\mbR_{j}.\nn
\eeq

We employ the regular expansion $\bW=\bar\bW+\ell\bW_{1}+\cdots$ and substitute it
into the resulting equation. 
The  leading-order equation 
\beq
\label{null}
\cP_{0}\bar \bW=0
\eeq
can be solved as follows \cite{RPK}.

For a positive (or negative) definite $\bL$, 
$\bL^{1/2}$ is well-defined and under the transformation
$\bL^{1/2}$ the matrix $  p_j\mbR_j\bL$ is transformed
into the Hermitian matrix $\bL^{{1/2}}  p_j\mbR_j\bL^{{1/2}}$
which has a complete set of eigenvectors and eigenvalues $\{\Om^\sigma\}\subset\IR$.
Let 
$\{\bd^{\sigma, \alpha}\}$ be the associated eigenvectors in
the original vector space   where the index $\alpha$ keeps track of the multiplicity. 
Let the eigenvectors $\{\bd^{\sigma,\alpha}\}$ 
be normalized such that $\bd^{\sigma,\alpha\dagger}\bL\bd^{\tau,\zeta}
=\delta_{\sigma,\tau}\delta_{\alpha,\zeta}$. 
It is easy to check that $\{\bb^{\sigma,\alpha\dagger}: \bb^{\sigma,\alpha}=\bL\bd^{\sigma,\alpha}\}$ 
are the {\em left} eigenvectors of $  p_j\mbR_j\bL$
and they are orthogonal to $\{\bd^{\tau,\zeta}(\bp)\}$ with respect
to the standard scalar product:
\beq
\label{lr}
\bb^{\sigma,\alpha\dagger}\bd^{\tau,\zeta}=\delta_{\sigma,\tau}\delta_{\alpha,\zeta}.
\eeq
Clearly, the eigenvalues $\Om^{\sigma}$ as a function of
the wavevector $\bp$ define the dispersion relations. 
For general  bianisotropic dielectric, it is easy to check that the zero eigenvalue $\Om^0=0$ is
always an eigenvalue with the associated left eigenvectors  
\beq
\label{zero}
\bb^{0,1}(\bp)\sim \lt(\begin{matrix}\bp\\
 0\end{matrix}\rt),\quad
\bb^{0,2}(\bp)\sim \lt(\begin{matrix}0\\
\bp\end{matrix}\rt). 
\eeq
Since
$\bL$ is invertible, it follows that
the null space of $  p_j\mbR_j\bL$ is spanned
by $\{\bd^{0,1}=\bK\bb^{0,1},\bd^{0,2}=\bK\bb^{0,2}\}$. 
The relations (\ref{lr}) and (\ref{zero}) imply that $\bd^{\tau,\zeta}, \tau\neq 0,$ are {\em transverse } vectors in the sense that
they are orthogonal to the wavevector $\bp$. 

Throughout the  English indices represent
the spatial degrees of freedom 
while the Greek indices represent the polarization
degrees of freedom. The Einstein summation convention and the Hermitian
conjugation are used only on the English indices.

Define 
\beq
\label{right}
\bDaz(\bp,\bq)&=&\bda(\bp)
\bdz(\bq)^\dagger\\
\label{left}
\bEaz(\bp,\bq)&=&\bba(\bp)
\bbz(\bq)^\dagger.
\eeq
The  null space of $\cP_{0}$ is 
the linear span of $\{\bD^{\tau,\alpha\zeta}(\bp,\bp),\forall \tau, \alpha,\zeta,\bp\}$, denoted by $\fM_\bp$,  for each $\bp\neq 0$ with the scalar product
$\hbox{Tr}\big[\bH^\dagger\bL\bG\bL\big],\bH,\bG\in \fM_\bp$. 

Then  the general
solution to (\ref{null}) can be expressed as 
\beq
\label{w0}
\bar \bW&=&\sum_{\sigma,\alpha,\zeta}\bar W^\sigma_{\alpha\zeta}\bDaz(\bp,\bp)
\eeq
where $\bar W^\sigma_{\alpha\zeta}$ are generally complex-valued functions.
The matrices $\bar \bW^\sigma=[\bar W^\sigma_{\alpha\zeta}]$, free of
the English indices, 
are referred to as the {\em coherence matrices}.

The constraint 
that the electric displacement $\bD$ and the magnetic induction 
$\bB$ 
are both divergence-free yields
\beqn
(\pm\nabla,\pm\nabla)\cdot\bar\bW=0
\eeqn
which,   in view of 
the definition (\ref{1.22}),  is equivalent to
\beq
(\pm\bp^{\dagger},\pm\bp^{\dagg})\bar \bW=0.
\eeq
Hence by (\ref{zero}) $\bb^{0,j\dagger} \bar\bW=0$
and by
 (\ref{lr})  $\bar\bW^0=0$ where $\bW^{0}$ is
 the $2\times 2$ coherence matrix associated with
 the non-propagating mode $\Om^{0}=0$. 
This implies, by (\ref{lr}), that $\bar\bW$ is a transverse field.

\commentout{
In contrary, when the same perturbation analysis is
applied to eq. (\ref{wig-mo2}) the $O(\ell^{-1})$-equation is
\beq
\label{null2}
  p_j\mbR_j\bL \bar \bW+\bar\bW\bL p_{j}\mbR_{j}=0
\eeq
whose solution is  entirely  different from  (\ref{w0})
and will be discussed in a separate publication. 
}

The $O(1)$ equation is
\beq
\label{1.13}
\cP_{0}\bW_{1}=i\om'\bar \bW
+\cP_{1}\bar\bW-\cP_{2}\bar\bW
\eeq
which is solvable if the right hand side is orthogonal to
the null space, $\fM_{\bp}$, of $\cP_{0}$.  The solvability
condition for (\ref{1.13}) then leads to the governing equation
for the $2\times 2$ coherence matrices:
\beq
\label{liou}
&&i\om'\bar \bW^{\tau}-\nabla_{\bx}\Om^{\tau}\cdot\nabla_{\bp} \bar \bW^{\tau}+\nabla_{\bp}\Om^{\tau}\cdot
\nabla_{\bx}\bar \bW^{\tau}
 -\bC^{\tau} \bar\bW^{\tau}-\bar\bW^{\tau}\bC^{\tau\dagger}=0
\eeq
where
the depolarization matrix $\bC^{\tau}=[C^{\tau}_{\xi\alpha}]$ is given by 
\beq
C^{\tau}_{\xi\alpha}=\partial_{x_{j}}\Om^{\tau}
\bb^{\tau,\xi\dagger}\partial_{p_{j}}\bd^{\tau,\alpha}
+\frac{1}{2}\lt[\partial_{x_{j}}\bb^{\tau,\xi\dagger}\mbR_{j} 
\bb^{\tau,\alpha} -\bb^{\tau,\xi\dagger}\mbR_{j}\pxj\bb^{\tau,\alpha}\rt].
\nn
\eeq
Using (\ref{lr}) we can cast $\bC^{\tau}$ in the
explicitly skew-symmetric (in $\xi, \alpha$) form:
\beq
C^{\tau}_{\xi\alpha}=\frac{1}{2}\partial_{x_{j}}\Om^{\tau}
\lt[\bb^{\tau,\xi\dagger}\bK\partial_{p_{j}}\bb^{\tau,\alpha}-\ppj\bb^{\tau,\xi\dagger}\bK\bb^{\tau,\alpha}\rt]
+\frac{1}{2}\lt[\partial_{x_{j}}\bb^{\tau,\xi\dagger}\mbR_{j} 
\bb^{\tau,\alpha} -\bb^{\tau,\xi\dagger}\mbR_{j}\pxj\bb^{\tau,\alpha}\rt].
\nn
\eeq
The details of the calculation is given in Appendix B.
Note that eq. (\ref{liou}) 
is invariant under the simultaneous  transformations of Hermitian conjugation and frequency exchange.

\section{Weak-disorder regime}
\label{sec3}
Now we consider the weak coupling
regime with the permittivity-permeability tensor $\bK$ 
given by 
\beq
\label{weak}
\bK^{-1}(\bx)=\bL_0(\bx)\lt[\bI+\sqrt{\ep} \bV\big(\frac{\bx}{\ep}\big)\rt],\quad\ell\ll 1
\eeq
where the  Hermitian matrix $\bL_0$
 represents the slowly varying  background medium
and  $\sqrt{\ep}\bV$ represents the medium fluctuations. 
The small parameter $\ep$ describes the ratio of
the scale  of the medium
fluctuation or the wavelength to the propagation distance
or the variability scale of $\bL_{0}$ .  

To preserve the Hermicity of $\bL$ the matrix  $\bV$ must satisfy
 \beq
 \label{herm}
 \bV^\dagger\bL_0=\bL_0\bV.
 \eeq
  We shall assume below that $\bL_0$ is either positive or negative definite. 
  A negative-definite $\bL_{0}$ gives rise to
negative  index of refraction \cite{SK, SPW}. 
A nondefinite $\bL_0$ gives rise  to complex-valued
refractive index and hence a lossy medium. 
To fix the idea, let us take $\bL_{0}$ to be positive definite.
Our method applies equally well to the negative definite case.

 % For the simplicity of presentation, we first
 % derive the results for 
  %the case of a $\bx$-independent $\bK_0$ and then 
  %extend it to the general case 
  %using  (\ref{go-final}). 

We assume that $\bV=[V_{ij}]$  is
a stationary (statistically homogeneous) random field with
the spectral density tensors ${\mb \Phi}=[\Phi_{ijmn}], {\mb \Psi}=[\Psi_{ijmn}]$ such that  
\beq
\lan V_{ij}(\bx)V^*_{mn}(\by)\ran&=&\int e^{i\bk^\dagger(\bx-\by)}\Phi_{ijmn}(\bk)d\bk\\
\lan V_{ij}(\bx)V_{mn}(\by)\ran&=&\int e^{i\bk^\dagger(\bx-\by)}\Psi_{ijmn}(\bk)d\bk
\eeq
which implies
\beq
\lan \hat V_{ij}(\bp)\hat V_{mn}^*(\bq)\ran&=&
\Phi_{ijmn}(\bp)\delta(\bp-\bq)\\
\lan \hat V_{ij}(\bp)\hat V_{mn}(\bq)\ran &=&
\Psi_{ijmn}(\bp)\delta(\bp+\bq). 
\eeq
Here and below $*$ denotes the complex conjugation. 
In the case of real-valued $\bV$, ${\mb \Phi}={\mb \Psi}$. 
The spectral density tensors  have 
the basic symmetry
\beq
\label{sym1}
\Phi_{ijmn}^*(\bp)&=&\Phi_{mnij}(\bp),\\
%\quad\forall i,j,m,n,\bp\\
\Psi_{ijmn}(-\bp)&=&\Psi_{mnij}(\bp),
%\quad\forall i,j,m,n.\bp
\label{sym2}
\eeq
Eq. (\ref{herm}) implies that
\beq
\label{15}
L_{0,ij}\Psi_{mnjl}(\bp)&=&
L^*_{0,lj}\Phi_{mnji}(\bp)\\
L_{0,ij}\Phi_{mnjl}(\bp)&=&L^*_{0,lj}\Psi_{mnji}(\bp)
%K_{0, ij}K_{0, mn}\Psi_{jlnr}(\bp)&=&K_{0,jl}K_{0,nr}\Psi^*_{nmji}(\bp).
\label{16}
\eeq
 %For real-valued $\bV$ the real-valuedness of ${\mb \Psi}$ is
%equivalent to the parity-invariance of  medium fluctuations. 

\beq
\nn
i\om'\bW
&=&\frac{i}{\ell}  p_j\mbR_j\int e^{i\bq^{\dagger}\bx}\widehat\bL_{0}(\bq) 
\bW(\bp-\frac{\ell\bq}{2})d\bq
-\frac{i}{\ell}\int \bW(\bp+\frac{\ell\bq}{2})
\widehat\bL_{0}(\bq)e^{i\bq^{\dagger}\bx}d\bq  p_j\mbR_j\\
&&+\frac{1}{2} \mbR_j\pxj
\int e^{i\bq^{\dagger}\bx}\widehat\bL_{0}(\bq)\bW(\bp-\frac{\ell\bq}{2})
d\bq+\frac{1}{2}\pxj\int
\bW(\bp+\frac{\ell\bq}{2})\widehat\bL_{0}(\bq)e^{i\bq^{\dagger}\bx}d\bq
\mbR_{j}\nn\\
&&+\frac{i}{\sqrt{\ell}}  p_j\mbR_j\int e^{i\bq^{\dagger}\tilde\bx}\widehat{\bL_{0}}(\frac{\bq'}{\ell})
\widehat{\bV}(\bq-\bq') 
\bW(\bp-\frac{\bq}{2})d\bq' d\bq\nn\\
&&
-\frac{i}{\sqrt{\ell}}\int \bW(\bp+\frac{\bq}{2})
\widehat{\big({\bV}^{\dagger}\big)}(\bq-\bq')\widehat{\bL_{0}}(\frac{\bq'}{\ell})e^{i\bq^{\dagger}\tilde\bx}d\bq' d\bq  p_j\mbR_j\nn\\
&&+\frac{1}{2\sqrt{\ell}} \mbR_j\pxjtil
\int e^{i\bq^{\dagger}\tilde\bx}\widehat {\bL_{0}}(\frac{\bq'}{\ell}) \widehat{\bV}(\bq-\bq')\bW(\bp-\frac{\bq}{2})
d\bq' d\bq\nn\\
&&+\frac{1}{2\sqrt{\ell}}\pxjtil\int
\bW(\bp+\frac{\bq}{2})\widehat{{\bV}^{\dagger}}(\bq-\bq')\widehat{\bL_{0}}(\frac{\bq'}{\ell})e^{i\bq^{\dagger}\tilde\bx}d\bq' d\bq
\mbR_j
\label{wig-eq'}\eeq
where $\tilde\bx=\bx/\ell$ is the fast spatial variable
and $\widehat{\big({\bV}^{\dagger}\big)}(\bq)=\widehat \bV^{\dagger}(-\bq)$ the Fourier transform of $\bV^{\dagger}$. 
As in the geometrical optics we approximate the first four terms
on the right hand side of eq. (\ref{wig-eq'}) by
$\ell^{-1}\cP_{0}\bW-\cP_{1}\bW+\cP_{2}\bW$.
For the last four terms on the right side of eq. (\ref{wig-eq'}) we have \beqn
\int e^{i\bq^{\dagger}\tilde\bx}\widehat{\bL_{0}}(\frac{\bq'}{\ell})
\widehat{\bV}(\bq-\bq') 
\bW(\bp-\frac{\bq}{2})d\bq' d\bq
&\approx& \bL_{0} (\bx)\int e^{i\bq^{\dagger}\tilde\bx}
\widehat{\bV}(\bq) 
\bW(\bp-\frac{\bq}{2}) d\bq\\
\int \bW(\bp+\frac{\bq}{2})
\widehat{{\bV}^{\dagger}}(\bq-\bq')\widehat{\bL_{0}}(\frac{\bq'}{\ell})e^{i\bq^{\dagger}\tilde\bx}d\bq' d\bq&\approx& \int \bW(\bp+\frac{\bq}{2})
\widehat {\bV}^{\dagger}(-\bq)e^{i\bq^{\dagger}\tilde\bx } d\bq \bL_{0}(\bx). 
\eeqn
Hence we have
the simplified form
\beq
\label{wig-eq}
i\om'\bW
&=& \ell^{-1}\cP_{0}\bW-\cP_{1}\bW+\cP_{2}\bW
+\ell^{-1/2}\cQ_{1}\bW+\ell^{-1/2}\cQ_{2}\bW
\eeq
where
\beq
\cQ_{1}\bW&=&{i}  p_j\mbR_j \bL_{0}\int e^{i\bq^{\dagger}\tilde\bx}\widehat\bV(\bq) 
\bW(\bp-\frac{\bq}{2})d\bq
-{i}\int \bW(\bp+\frac{\bq}{2})
\widehat\bV^{\dagger}(\bq)e^{i\bq^{\dagger}\tilde\bx}d\bq \bL_{0}  p_j\mbR_j\nn\\
\cQ_{2}\bW&=&\frac{1}{2} \mbR_j\bL_{0}\pxjtil
\int e^{i\bq^{\dagger}\tilde\bx}\widehat\bV(\bq)\bW(\bp-\frac{\bq}{2})
d\bq+\frac{1}{2}\pxjtil\int
\bW(\bp+\frac{\bq}{2})\widehat\bV^{\dagger}(-\bq)e^{i\bq^{\dagger}\tilde\bx}d\bq
\bL_{0}\mbR_{j}\nn
\eeq
Hereafter we shall work with eq. (\ref{wig-eq}) to derive the
2f-RT equations by emplying the multiscale expansion (MSE) . 

\section{Multiscale expansion}
\label{sec4}
The key point of MSE is to separate the fast variable $\tilde\bx$ from
the slow variable $\bx$ and make the substitution 
\beqn
\nabla\bW&\to &
\nabla_{\bx}\bW+
\ep^{-1}\nabla_{\tilde \bx}\bW. 
\eeqn
Consequently, 
\beq
\cP_{2}\bW&\rightarrow & \cP_{2}\bW+\ell^{-1}\widetilde\cP_{2}\bW\nn
\eeq
with 
\beq
\cP_{2}\bW&=&\frac{1}{2} \mbR_j\pxj
\lt[\bL\bW\rt]+\frac{1}{2}\pxj
\lt[\bW \bL\rt]
\mbR_j\label{29}\\
\widetilde\cP_{2}\bW&=&\frac{1}{2} \mbR_{j}\bL\pxjtil
\bW +\frac{1}{2}\pxjtil
\bW \bL \mbR_{j}.\label{30}
\eeq
The idea is that for sufficiently small $\ep$ the two widely separated scales, represented by $\bx $ and $\tilde\bx$ respectively, become  mathematically independent.

We posit the expansion
$\bW=\bar \bW+\sqrt{\ep} \bW_1+\ep \bW_2+...$, substitute  it into eq. (\ref{wig-eq}) and equate 
terms of same order of magnitude.

The $O(\ep^{-1})$ equation is 
\beq
\label{eq1}
(\widetilde\cP_{2}+\cP_{0})\bar\bW=0. 
%=\mbR_j\bL_{0}\frac{\partial}{\partial \tilde x_j}\bar\bW+\frac{\partial}{\partial \tilde x_j} \bar \bW\bL_{0}\mbR_j+{2i}\lt[p_j\mbR_j\bL_{0}\bar \bW-\bar \bW \bL_{0}  p_j\mbR_j\rt]=0. 
\eeq
We hypothesize that the leading order term $\bar\bW=\bar\bW(\bx,\bp)$ be
independent of the fast variable $\tilde\bx$. Thus
$\widetilde\cP_{2}\bar\bW=0$ and 
eq. (\ref{eq1})  reduces to
 (\ref{null}) and 
its solution takes the form (\ref{w0}). 

The $O(\ep^{-1/2})$-equation is
\beq
\label{eq3}
\cP_{0}\bW_{1}+\widetilde\cP_{2}\bW_{1}=-\cQ_{1}\bar\bW-\cQ_{2}\bar\bW
\eeq
or equivalently, after Fourier-transforming-in-$\tilde\bx$ and adding a regularizing $O(\ell)$-term
\commentout{
\beq
\label{eq3}
&&\ep\bW_1+\frac{1}{2}\mbR_j\bL_{0}\frac{\partial}{\partial \tilde x_j}\bW_1+\frac{1}{2}\frac{\partial}{\partial \tilde x_j} \bW_1\bL_{0}\mbR_j+{i}p_j\mbR_j \bL_{0}\bW_1- i\bW_1 \bL_{0} p_j\mbR_j\\
&=&
-\frac{i}{2}\mbR_j\bL_{0}
\int (p_{j}+\frac{q_{j}}{2}) e^{i\bq^{\dagger}\tilde\bx}\widehat\bV(\bq)\bW(\bp-\frac{\bq}{2})
d\bq+\frac{i}{2}\int
\bW(\bp+\frac{\bq}{2})\widehat\bV^{\dagger}(-\bq)e^{i\bq^{\dagger}\tilde\bx}(p_{j}-\frac{q_{j}}{2})d\bq
\bL_{0}\mbR_{j}. \nn
\eeq
We  Fourier transform eq. (\ref{eq3}) in $\tilde\bx$
}
\beq
\nn
&&-i2\ep \widehat \bW_1(\bk, \bp)+ k_j\mbR_j\bL_{0}\widehat\bW_1(\bk, \bp)+\widehat\bW_1(\bk, \bp) \bL_{0}k_j\mbR_j\nn\\
&&+\nn
{2}\lt[  p_j\mbR_j\bL_{0}\widehat \bW_1(\bk, \bp)-\widehat\bW_1(\bk,\bp)\bL_{0}
  p_j\mbR_j\rt]\\
&&=-\mbR_{j}\bL_{0}(2p_{j}+{k_{j}})\widehat\bV(\bk) \bar\bW(\bp-\frac{\bk}{2})
+\bar\bW(\bp+\frac{\bk}{2})\widehat\bV^\dagger(-\bk)
(2p_{j}-{k_{j}})\bL_{0}\mbR_{j}
\label{eq4}
\eeq
and posit
the solution 
\beq
\label{w1}
\widehat \bW_1(\bk,\bp)=\sum_{\sigma,\alpha,\zeta}\Caz(\bk, \bp)
\bDaz\big(\bp+\frac{\bk}{2},\bp-\frac{\bk}{2}\big)
\eeq
where $\Caz$ are generally complex numbers. 
Note that the two arguments of $\bDaz$ in (\ref{w1}) are at different
momenta $\bp+{\bk}/{2},\bp-{\bk}/{2}$. 
%For ease of notation, we will omit writing
%the argument $\bx$ below.

We substitute (\ref{w0}) and (\ref{w1}) into eq. (\ref{eq4})
and multiply it with $\bba(\bp+\frac{\bk}{2})^\dagger$ from
the left and with $\bbz(\bp-\frac{\bk}{2})$ from the right
and solve the resulting equation algebraically. This
 yields the coefficients 
\beq
\nn
\Caz(\bk, \bp)&=&\lt(\Om^\sigma(\bp+\frac{\bk}{2})
-\Om^\sigma(\bp-\frac{\bk}{2})-i\ep\rt)^{-1}\sum_{\eta}\nn\\
&&\lt[-\Om^{\sigma}(\bp+\frac{\bk}{2})\bar W^\sigma_{\eta\zeta}(\bp-\frac{\bk}{2})\bba(\bp+\frac{\bk}{2})^\dagger\widehat\bV(\bk)\bde(\bp-\frac{\bk}{2})\rt.\nn\\
&&\lt. +\Om^{\sigma}(\bp-\frac{\bk}{2})\bar W^\sigma_{\alpha\eta}(\bp+\frac{\bk}{2})\bde(\bp+\frac{\bk}{2})^\dagger
\widehat\bV^\dagger(-\bk)
\bbz(\bp-\frac{\bk}{2})\rt].\label{cij}
\eeq
When  the leading term $\bar \bW$ is invariant
under the simultaneous  transformations of Hermitian conjugation $\dagger$ and frequency exchange $\omega_1\leftrightarrow \omega_2$,
so is $\bW_{1}$ which is equivalent to
 \[
 C^{\sigma*}_{\zeta\alpha}(-\bk,\bp; \omega_1, \omega_2)
 =\Caz(\bk, \bp;\omega_2, \omega_1). 
 \]

Finally the $O(1)$-terms yields the equation after regularization
\beq
\ep\bW_2+\frac{1}{2}\mbR_j\bL_{0}\frac{\partial}{\partial \tilde x_j}\bW_2+\frac{1}{2}\frac{\partial}{\partial \tilde x_j} \bW_2\bL_{0}\mbR_j+{i}p_j\mbR_j\bL_{0}\bW_2- i\bW_2  \bL_{0}p_j\mbR_j=\bF\label{eq5}
\eeq
with
\beq
\bF&=&i\om'\bar \bW+\cP_{1}\bar\bW-\cP_{2}\bar\bW-\cQ_{1}\bW_{1}-\cQ_{2}\bW_{1}
 \label{F}. 
\eeq

It suffices to note 
that in order for  the resulting solution $\ep\bW_2$  
to vanish in the limit $\ep\to 0$, $\bF$
must satisfy the solvability condition
\beq
\label{sol}
\lim_{\ell\to 0}\hbox{Tr}\lan \bG^\dagger\bL_0\bF\bL_0\ran=0
\eeq
for all random stationary matrices $\bG$ satisfying eq. (\ref{eq1}). 
This can be seen  by transforming eq. (\ref{eq5}) 
into $\hbox{Tr} \lan \bG^\dagger \bL_0 (\ref{eq5})\bL_0\ran$ which by eq. (\ref{eq1}) implies
$2\ell\hbox{Tr} \lan \bG^\dagger \bL_0 \bW_2\bL_0\ran
=\hbox{Tr} \lan \bG^\dagger\bL_0\bF\bL_0\ran$
and hence (\ref{sol}). 
 \commentout{This follows from
the identity
\beqn
&&\lim_{\ell\to 0}\hbox{Tr}\lan \bG^\dagger\bK_0\bF\bK_0\ran\\
&=&-\lim_{\ell\to 0}\hbox{Tr}\lan\lt(\bK_0^{-1}\mbR_j\frac{\partial}{\partial \tilde x_j}\bG+\frac{\partial}{\partial \tilde x_j} \bG\mbR_j\bK_0^{-1}+2i\lt[\bK_0^{-1}  p_j\mbR_j\bG- \bG  p_j\mbR_j\bK_0^{-1}\rt]\rt)\bK_0\bW^\ell_2\bK_0\ran
\eeqn
derived from the left hand side of eq. (\ref{eq5})
by integrating by parts and using the cyclic property of
trace operation. 
}

Fortunately, we do not need to work with the full solvability condition (\ref{sol}). It suffices to demand (\ref{sol}) to be fulfilled
by all {\em deterministic} $\bG$, independent
of $\tilde\bx$, such that
\beq
  p_j\mbR_j\bL_{0}\bG- \bG \bL_{0} p_j\mbR_j=0.
\eeq
In other words, as in (\ref{w0}),  we consider  only a subspace
of the null space of eq. (\ref{eq1}) and  replace (\ref{sol})  by 
\beq
\label{fred}
\lim_{\ell\to 0}\rm{Tr}\lt(\bE^{\tau, \xi\nu\dagger}(\bp,\bp)
\lan \bF(\bx,\tilde\bx,\bp)\ran\rt)=0,\quad\forall \tau, \xi,\nu,\bx,\tilde\bx,\bp
\eeq
where $\bE^{\tau,\xi\nu}$ are defined in (\ref{left}). 
 As  (\ref{wig-eq}), (\ref{eq3})
and (\ref{F}) are invariant under 
the simultaneous  transformations of Hermitian conjugation $\dagger$ and frequency exchange $\omega_1\leftrightarrow \omega_2$, therefore
eq. (\ref{fred}) must also be invariant under the same transformations.

To summarize, we have constructed the asymptotic solution $\bar\bW+\sqrt{\ell}\bW_{1}+\ell \bW_2$ which 
satisfies approximately 
 the 2f Wigner-Moyal equation
 in
the sense that the remainder vanishes in a suitable sense as $\ell\to 0$ \cite{2frt-helm, 2frt-maxwell}.

With (\ref{w1})-(\ref{cij}) and (\ref{F}), eq. (\ref{fred})
is an implicit form of 2f-RT equations  that determines
the leading order coherence matrix.  Our next step
is to write (\ref{fred}) explicitly in terms of explicit, physical quantities. 

\section{2f-RT equations}
\label{sec5}
As in the geometrical optics, 
the terms $i\om'\bar \bW+\cP_{1}\bar\bW-\cP_{2}\bar\bW$ in the expression (\ref{F}) yield
the left hand side of (\ref{liou}) after the operation (\ref{fred}). 

First note the key expression
\beq
&&{\lan \int d\bq e^{i\bq\dagger\tilde\bx} \hat \bV(\bq)
\bW_{1}(\bp-\frac{\bq}{2}) \ran_{sj}\nn}\\
&=&
\sum_{\sigma, \alpha, \zeta, \eta}
\int d\bk 
\lt(\Om^\sigma(\bp+\bk)-\Om^\sigma(\bp)-i\ell\rt)^{-1}\nn\\
&&\times\Big[-\Om^{\sigma}(\bp+\bk)\bar W^\sigma_{\eta\zeta}(\bp) 
e^{\sigma,\alpha*}_f(\bp+\bk)\Psi_{fgsi}(\bk)
d^{\sigma,\eta}_g(\bp)
D^{\sigma, \alpha\zeta}_{ij}(\bp+\bk,\bp)\nn\\
&&+\Om^{\sigma}(\bp)\bar W^{\sigma}_{\alpha\eta}(\bp+\bk)
d^{\sigma,\eta*}_{g}(\bp+\bk)\Phi^{*}_{fgsi}(-\bk)e_{f}^{\sigma,\zeta}(\bp)D^{\sigma,\alpha\zeta}(\bp+\bk,\bp)\Big]\nn
\eeq
and
\beq
&&\lan \int d\bq e^{i\bq\dagger\tilde\bx} 
\bW_{1}(\bp+\frac{\bq}{2}) \hat \bV(-\bq)^{\dagger}\ran_{sj}\nn\\
&=&
\sum_{\sigma, \alpha, \zeta, \eta}
\int d\bk 
\lt(\Om^\sigma(\bp)-\Om^\sigma(\bp-\bk)-i\ell\rt)^{-1}\nn\\
&&\times\Big[-\Om^{\sigma}(\bp)\bar W^\sigma_{\eta\zeta}(\bp-\bk)
e^{\sigma,\alpha}_f(\bp)^*\Phi_{fgjn}(\bk)
 d_{g}^{\sigma,\eta}(\bp-\bk)
D^{\sigma, \alpha\zeta}_{sn}(\bp, \bp-\bk)\nn\\
&&+\Om^{\sigma}(\bp-\bk)\bar W^{\sigma}_{\alpha\eta}(\bp)
d^{\sigma,\eta*}_{g}(\bp)\Psi^{*}_{fgjn}(-\bk)e_{f}^{\sigma,\zeta}(\bp-\bk)D^{\sigma,\alpha\zeta}_{sn}(\bp,\bp-\bk)\Big]. \nn
\eeq
The above expressions are independent of the fast variable $\tilde\bx$ so $\lan \cQ_{2}\bW_{1}\ran=0$. 
We also have
\beqn
&&\rm{Tr}\lt(\bE^{\sigma,\xi\nu\dagger}(\bp,\bp)
\lan \cQ_{1}\bW_{1}\ran\rt)\\
&=&i\sum_{\alpha,\eta}\int d\bk \lt(\Om^\tau(\bp+\bk)-\Om^\tau(\bp)-i\ell\rt)^{-1}
\Om^{\sigma}(\bp)e^{\sigma,\xi*}_{s}(\bp)\nn\\
&&\times\Big[-\Om^{\sigma}(\bp+\bk)\bar W^\sigma_{\eta\nu}(\bp) 
e^{\sigma,\alpha*}_f(\bp+\bk)\Psi_{fgsi}(\bk)
d^{\sigma,\eta}_g(\bp)
d^{\sigma, \alpha}_{i}(\bp+\bk)\nn\\
&&+\Om^{\sigma}(\bp)\bar W^{\sigma}_{\alpha\eta}(\bp+\bk)
d^{\sigma,\eta*}_{g}(\bp+\bk)\Phi^{*}_{fgsi}(-\bk)e_{f}^{\sigma,\nu}(\bp)d_{i}^{\sigma,\alpha}(\bp+\bk)\Big]\\
&&-i
\sum_{ \zeta, \eta}
\int d\bk 
\lt(\Om^\sigma(\bp)-\Om^\sigma(\bp-\bk)-i\ell\rt)^{-1}\Om^{\sigma}(\bp) e^{\sigma,\nu}_{j}(\bp)\nn\\
&&\times\Big[-\Om^{\sigma}(\bp)\bar W^\sigma_{\eta\zeta}(\bp-\bk)
e^{\sigma,\xi*}_f(\bp)\Phi_{fgjn}(\bk)
 d_{g}^{\sigma,\eta}(\bp-\bk)
d^{\sigma,\zeta*}_{n}(\bp-\bk)\nn\\
&&+\Om^{\sigma}(\bp-\bk)\bar W^{\sigma}_{\xi\eta}(\bp)
d^{\sigma,\eta*}_{g}(\bp)\Psi^{*}_{fgjn}(-\bk)e_{f}^{\sigma,\zeta}(\bp-\bk)d^{\sigma,\zeta*}_{n}(\bp-\bk)\Big].
\eeqn

To state the full result in a concise form, let
us introduce the following quantities. 
Define the  scattering kernel tensors ${\fS}^{\tau}(\bp,\bq)=[\cK^{\tau}_{\xi\nu\alpha\zeta}(\bp,\bq)]$ as
\beq
\label{kernel}
\cK^{\tau}_{\xi\alpha \nu\zeta }(\bp,\bq)&= &\Om^{\tau}(\bp)\Om^{\tau}(\bq)e^{\tau,\xi*}_s(\bp) d^{\tau,\alpha}_i\big(\bq\big) 
\Phi_{sifg}\big(\bp-\bq\big)e^{\tau, \nu}_f\big(\bp\big)d^{\tau,\zeta*}_g\big(\bq\big)
\eeq
Using (\ref{sym1})-(\ref{16}) one can derive the alternative
expressions  in terms of $\mathbf{\Psi}$
\beqn
\cK^{\tau}_{\xi\alpha\nu\zeta}(\bp,\bq) 
 &=&\Om^{\tau}(\bp)\Om^{\tau}(\bq)e^{\tau,\nu}_f(\bp)  d^{\tau, \zeta*}_{g}(\bq)\Psi^{*}_{fgsi}\big(\bq-\bp\big)
 e^{\tau,\alpha}_s(\bq) d^{\tau,\xi*}_i(\bp)
\nn\\
&=&\Om^{\tau}(\bp)\Om^{\tau}(\bq)e^{\tau,\xi*}_s(\bp)
d^{\tau,\alpha}_i(\bq)\Psi_{sifg}\big(\bq-\bp\big)e^{\tau,\zeta*}_f(\bq)d^{\tau, \nu}_{g}(\bp)
\eeqn
and the properties
\beq
\label{adj1}
\cK^{\tau}_{\xi\alpha\nu\zeta}(\bp,\bq)&=&\cK^{\tau*}_{\nu\zeta\xi\alpha}(\bp,\bq)
\\
\cK^{\tau}_{\xi\alpha\nu\zeta}(\bq,\bp)&=&\cK^{\tau}_{\zeta\nu\alpha\xi}(\bp,\bq)\label{adj2}
\eeq

For any $\fM_\bp$-valued field $\bG(\bp)$ define
the $(\xi,\nu)$-component of the tensor
$\fS^{\tau}(\bp,\bq):\bG(\bq)$ as
\[
\lt[\fS^{\tau}(\bp,\bq):\bG(\bq)\rt]_{\xi\nu}=\sum_{\alpha,\zeta}\cK^{\tau}_{\xi\alpha\nu\zeta}(\bp,\bq)
G_{\alpha\zeta}(\bq).
\]
Define the tensors $\bSig^\tau=[\Sigma^\tau_{\xi\nu}]$ analogous to the total scattering cross section 
  as
\beq
\label{sig}
\bSig^{\tau}(\bp)&= &\lt\{\int
\delta\Big( \Om^\tau(\bp)-\Om^\tau(\bq)\Big)- i\int\cpv \lt(\Om^\tau(\bp)-\Om^\tau(\bq)\rt)^{-1}\rt\} \fS^{\tau}
(\bp,\bq):\bI d\bq.\nn
%&&\sum_{\alpha} d^{\tau,\xi*}_s(\bp)  d^{\tau,\nu}_f(\bp)
%\Phi_{sife}\big(k(\bp-\bq)\big)e^{\tau,\alpha*}_e(\bq)
%e^{\tau, \alpha}_{i}(\bq)d\bq.\label{total}
\eeq
%For a Hermitian $\bV$, we have $(\bSig^-)^\dagger
%=\bSig^+$. 
The  2f-RT equations for the coherence matrix $\bar\bW^{\tau}$ then reads as
\beq
\label{2frt}
&&i\om'\bar \bW^{\tau}-\nabla_{\bx}\Om^{\tau}\cdot\nabla_{\bp} \bar \bW^{\tau}+\nabla_{\bp}\Om^{\tau}\cdot
\nabla_{\bx}\bar \bW^{\tau}
 -\bC^{\tau} \bar\bW^{\tau}-\bar\bW^{\tau}\bC^{\tau\dagger} \\
&=&2\pi\int \delta\Big(\Om^\tau(\bq)-\Om^\tau(\bp)\Big)
\fS^{\tau}(\bp,\bq):\bar \bW^\tau(\bq) d\bq-\bSig^{\tau}(\bp)\bar\bW^\tau(\bp)-\bar\bW^\tau(\bp)\bSig^{\tau\dagger}(\bp),\quad \forall\tau.\nn
\eeq
The $\delta$-function and Cauchy singular kernel arise because of the
fact
\beq
\label{cpv}
\lim_{\ell\to 0}\frac{1}{x -i\ell}=i\pi \delta(x)+\frac{1}{x}
\eeq
in the sense of generalized function. 

With the property (\ref{adj1}) one can verify directly
the invariance of (\ref{2frt}) with respect to 
the simultaneous  transformations of Hermitian conjugation and frequency exchange.

 \subsection{Birefringence: scalar 2f-RT equation}
\label{sec:scalar}
Although,   in view of (\ref{zero}), the zero eigenvalue $\Om^0=0$ has multiplicity
two in general, the nonzero eigenvalues in media other than
the simplest isotropic medium often have multiplicity one
as we shall see in Section~\ref{sec:med}. This is closely
related to the birefringence effect. Under such
circumstances, the 2f-RT equations take a much
simplified form which we now state. 

 Because $\Omega^{j}, j=1,2,3,4$ are simple (of multiplicity one), expression (\ref{w0}) reduces to 
 \beqn
\bar \bW&=&\sum_{\sigma}\bar W^\sigma\bD^{\sigma}(\bp,\bp).
\eeqn
Consequently (\ref{2frt}) becomes
a scalar equation for $\bar W^{\sigma}$ and
the different polarization modes decouple:
\beq
\label{2frt2}
&&i\om'\bar W^{\tau}-\nabla_{\bx}\Om^{\tau}\cdot\nabla_{\bp} \bar W^{\tau}+\nabla_{\bp}\Om^{\tau}\cdot
\nabla_{\bx}\bar W^{\tau}\\
&=&{2\pi}\int \delta\Big(\Om^\tau(\bq)-\Om^\tau(\bp)\Big)
\fS^\tau(\bp,\bq)\bar W^\tau(\bq) d\bq\nn-2\Sigma^\tau(\bp)\bar W^\tau(\bp),\quad \forall\tau\nn
\eeq
where 
\beq
\label{kernel-scalar}
%C^{\tau}&=&\partial_{x_{j}}\Om^{\tau}
%\bb^{\tau\dagger}\bK_{0}\partial_{p_{j}}\bb^{\tau}
%+\frac{1}{2}\lt[\partial_{x_{j}}\bb^{\tau\dagger}\mbR_{j} 
%\bb^{\tau} -\bb^{\tau\dagger}\mbR_{j}\pxj\bb^{\tau}\rt]
%\nn\\
\fS^\tau(\bp,\bq)&=&\Om^\tau(\bp)\Om^\tau(\bq)e^{\tau*}_s(\bp) d^{\tau}_{i}\big(\bq\big)  
\Phi_{sifg}\big(\bp-\bq\big) e^{\tau}_f\big(\bp\big)d^{\tau*}_g\big(\bq\big)\\
\Sigma^\tau(\bp)
&=&\pi\int
\delta\Big( \Om^\tau(\bp)-\Om^\tau(\bq)\Big) \fS^\tau
(\bp,\bq) d\bq.
\label{tot-scalar}
\eeq
Note that the Cauchy principal value integral disappears
from (\ref{tot-scalar}) whenever
$\Sigma^\tau$ and $\bar\bW$ commute.

\subsection{Paraxial approximation: spatial anisotropy}
\label{sec8}

Consider now  a {\em spatially}  anisotropic spectral density
tensor
for  a medium fluctuating much more
slowly in the longitudinal $x_3=z$ direction, i.e.
replacing $\mathbf{\Phi}\big(\bp-\bq\big)$ in (\ref{2frt})  by 
\[
\frac{1}{\theta}\mathbf{\Phi}\lt( \bp_\perp-\bq_\perp, \frac{1}{\theta}(p-q)\rt)
\]
which, in the limit $\theta\to 0$, tends to
\beq
\label{aniso}
\delta(p-q) \int dk \mathbf{\Phi}\lt(
\bp_\perp-\bq_\perp, k\rt). 
\eeq
With
$ \bar \bW^\sigma= \bar \bW^\sigma(\bx_\perp,z, \bp_\perp, p)$,
the right hand side of eq. (\ref{2frt}) reduces to
\beqn
&&2\pi\int \delta\Big(\Om^\tau(\bp_\perp, p)-\Om^\tau(\bq_\perp, p)\Big)
\fS(\bp_\perp,\bq_\perp):\bar\bW^{\tau}(\bq_\perp) d\bq_{\perp}\nn\\
&&-\Big[\bSig^\tau(\bp_\perp)\bar\bW^\tau(\bp_\perp)+\bar\bW^\tau(\bp_\perp)\bSig^{\tau\dagger}(\bp_\perp)\Big]
\eeqn 
where $\fS^\tau=[\cS_{\xi\alpha\nu\zeta}^\tau]$,
\beqn
\label{kernel-para}
\cK^{\tau}_{\xi\alpha\nu\zeta}(\bp_\perp,\bq_\perp)&= &e^{\tau,\xi*}_s(\bp_\perp,p) d^{\tau, \alpha}_{i}\big(\bq_\perp,p\big)
\int \Phi_{sifg}\big(\bp_\perp-\bq_\perp, k\big){d k }e^{\tau,\nu}_f\big(\bp_\perp,p\big)  d^{\tau,\zeta*}_g\big(\bq_\perp,p\big)\eeqn
and 
\beqn
\bSig^\tau(\bp_\perp)
&=&\lt[\int
\delta\Big( \Om^\tau(\bp_\perp, p)-\Om^\tau(\bq_\perp, p)\Big) - i\int\cpv
\Big( \Om^\tau(\bp_\perp, p)-\Om^\tau(\bq_\perp, p)\Big)^{-1}\rt]\fS^\tau
(\bp_\perp,\bq_\perp):\bI d\bq_{\perp}. 
\eeqn
 Eq. (\ref{2frt}) now takes
the paraxial form
\beq
\label{para}
&&\partial_{p}\Om^{\tau}\partial_z \fW^\tau+\nabla_{\bp_{\perp}}\Om^{\tau}\cdot
\nabla_{\bx_{\perp}}\bar \bW^{\tau}+i\om'\bar \bW^{\tau}-\nabla_{\bx}\Om^{\tau}\cdot\nabla_{\bp} \bar \bW^{\tau}
 -\bC^{\tau} \bar\bW^{\tau}-\bar\bW^{\tau}\bC^{\tau\dagger} 
\nn\\
&=&2\int \delta\Big(\Om^\tau(\bp_\perp, p)-\Om^\tau(\bq_\perp, p)\Big)
\fS(\bp_\perp,\bq_\perp):\fW^\tau(\bq_\perp) d\bq_{\perp}\nn\\
&&-\Big[\bSig^\tau(\bp_\perp)\fW^\tau(\bp_\perp)+\fW^\tau(\bp_\perp)\bSig^{\tau\dagger}(\bp_\perp)\Big].
\eeq
The longitudinal variable $z$ plays the role of a temporal variable
and $p$ is a parameter so that 
(\ref{para})  can be solved as an ``initial'' value problem given
the initial  data on $z=\hbox{constant}$
and a fixed $p$ if $\partial_{p}\Om^{\tau}\neq 0$.

\section{Examples}
In this section, we briefly discuss a few media
for which the scattering tensor can be explicitly computed
(see \cite{2frt-maxwell} for a more elaborate discussion).

\label{sec:med}
\subsection{Isotropic medium}
\label{sec:iso}
\label{sec6}
For the simplest isotropic medium, $\bK_{0}=\hbox{diag}\big[\epsilon_{0},\epsilon_{0},\epsilon_{0},\mu_{0},\mu_{0},\mu_{0}\big]$.
There are two nozero eigenvalues:
$ \Om^+(\bp)=c_{0}|\bp|, \Om^-(\bp)=-c_{0}|\bp|$ of
multiplicity two. Let $\hat \bp=\bp/|\bp|$ and let $\hat\bp^+_\perp, \hat\bp^-_\perp$ be any pair of unit vectors orthogonal to each other and to $\hat\bp$ so that $\{\hat\bp^+_\perp, \hat\bp^-_\perp,\hat\bp\}$ form a right-handed coordinate frame. Let $\{\hat\bq^+_\perp, \hat\bq^-_\perp,\hat\bq\}$ 
be similarly defined. The 
eigenvectors are
\beqn
%\label{b1}
%\bd^{0,1}(\bp)={\sqrt{\epsilon_0}}(\hat\bp, 0)^\dagger,& \bd^{0,2}(\bp)={\sqrt{\epsilon_0}}(0,\hat\bp)^\dagger,&\hbox{for}\quad\Om^0(\bp);\\
\bd^{+, +}(\bp)=\lt(\begin{matrix}
\sqrt{\frac{\epsilon_{0}}{{2}}}\hat\bp^+_\perp\\
\sqrt{\frac{\mu_{0}}{2}}\hat\bp^-_\perp\end{matrix}\rt),
\bd^{+, -}(\bp)=\lt(\begin{matrix}\sqrt{\frac{\epsilon_{0}}{2}}\hat\bp^-_\perp\\
 -\sqrt{\frac{\mu_{0}}{2}}\hat\bp^+_\perp\end{matrix}\rt),
\bd^{-, +}(\bp)=\lt(\begin{matrix}\sqrt{\frac{\epsilon_{0}}{2}}\hat\bp^+_\perp\\
 -\sqrt{\frac{\mu_{0}}{2}}\hat\bp^-_\perp\end{matrix}\rt),
\bd^{-, -}(\bp)=\lt(\begin{matrix}\sqrt{\frac{\epsilon_{0}}{2}}\hat\bp^-_\perp\\
 \sqrt{\frac{\mu_{0}}{2}}\hat\bp^+_\perp\end{matrix}\rt).
\eeqn

Often, in a scattering atmosphere for instance, $\tilde\mu\approx 0$ and consequently 
\beq
\label{chan}
\cK^\tau_{\xi\alpha\nu\zeta}(\bp,\bq)&=&
\frac{1}{4}\Phi_\epsilon(\bp-\bq)\hat\bp^{\xi\dagger}_\perp\hat\bq^{\alpha}_\perp\hat\bq^{\zeta\dagger}_\perp\hat\bp^{\nu}_\perp,\quad \tau,\xi,\alpha,\nu, \zeta=\pm.
\eeq
This is the  setting
for which S. Chandrasekhar originally derived 
his famous equation of transfer \cite{Cha}. 
In this case, eq. (\ref{2frt}) is the two-frequency version of
Chandrasekhar's famous transfer equation \cite{Cha, Han}.

As we shall see below, many materials are birefringent
and  permit
 two monochromatic plane waves with two different
 linear polarizations and two different 
 velocities to propagate in any given
 direction \cite{BW}. This is the birefringence effect.

\subsection{Chiral media}
\label{sec:chi}
A chiral medium is a reciprocal, biisotropic medium with
 the constitutive matrix 
\[
\bK_{0}=\lt[\begin{matrix}
\epsilon_{0} \bI&i\chi \bI\\
-i\chi\bI& \mu_{0} \bI
\end{matrix}\rt]
\]
where $\chi\in \IR$ is the magneto-electric coefficient.
To maintain a positive-definite $\bK_{0}  $ we assume  $\chi^{2}<\epsilon\mu$. We then have
\beq
\label{mat}
 p_j\mbR_j\bL_{0}=
\frac{c_{0}}{1-\kappa^{2}}
\lt[\begin{matrix}
0&-\bp\times\\
\bp\times &0
\end{matrix}
\rt]\lt[\begin{matrix}
z\bI&-i\kappa\bI\\
i\kappa\bI&z^{{-1}}\bI
\end{matrix}\rt]
\eeq
\commentout{
\[
 A=\frac{1}{(1-\kappa^{2})^{6}}
\left[ \begin{array}{cccccc}
z&0&0&-i\kappa&0&0\\
0&z&0&0&-i\kappa&0\\
0&0&z&0&0&-i\kappa\\
i\kappa&0&0&1/z&0&0\\
0&i\kappa&0&0&1/z&0\\
0&0&i\kappa&0&0&1/z
\end{array}\right]
\left[\begin{array}{cccccc}
0&0&0&0&p_{3}&-p_{2}\\
0&0&0&-p_{3}&0&p_{1}\\
0&0&0&p_{2}&-p_{1}\\
0&-p_{3}&p_{2}&0&0&0\\
p_{3}&0&-p_{1}&0&0&0\\
-p_{2}&p_{1}&0&0&0&0
\end{array}
\right].
\]
}
where $z=\sqrt{\mu_{0}/\epsilon_{0}}>0$ is the impedance and
$ \kappa=\chi c_{0}$ is the chirality parameter.
The  four non-zero simple eigenvalues 
and their  corresponding 
eigenvectors are
\beqn
\bb^{1}&\sim \lt(\begin{matrix}
-i\hat\bp^{1}_{\perp}+\hat\bp^{2}_{\perp}\\
-z^{-1}\hat\bp^{1}_{\perp}-iz^{-1}\hat\bp^{2}_{\perp}
\end{matrix}\rt), &\Omega^{1}=c_{0}|\bp|(1+\kappa)^{-1};\\
\bb^{2}&\sim  \lt(\begin{matrix}
i\hat\bp^{1}_{\perp}+\hat\bp^{2}_{\perp}\\
-z^{-1}\hat\bp^{1}_{\perp}+iz^{-1}\hat\bp^{2}_{\perp}
\end{matrix}\rt), & \Omega^{2}=c_{0}|\bp|(1-\kappa)^{-1};\\
\bb^{3}&\sim \lt(\begin{matrix}
-i\hat\bp^{1}_{\perp}+\hat\bp^{2}_{\perp}\\
z^{-1}\hat\bp^{1}_{\perp}+iz^{-1}\hat\bp^{2}_{\perp}
\end{matrix}\rt), & \Omega^{3}=c_{0}|\bp|(\kappa-1)^{-1};\\
\bb^{4}&\sim \lt(\begin{matrix}
i\hat\bp^{1}_{\perp}+\hat\bp^{2}_{\perp}\\
z^{-1}\hat\bp^{1}_{\perp}-iz^{-1}\hat\bp^{2}_{\perp}
\end{matrix}\rt), & \Omega^{4}=c_{0}|\bp|(-\kappa-1)^{-1}. 
\eeqn
Note also that $\Om^{4}=-\Om^{1}, \Om^{3}=-\Om^{2}$. 
As $|\kappa|<1$, $\bb^{1}, \bb^{2}$ are the forward propagating modes and $\bb^{3},\bb^{4}$ the backward
propagating modes. 

 \subsection{Birefrigence in anisotropic crystals}
 \label{sec:aniso}
  The only optically isotropic crystal is the
 cubic crystal. 
 In the system of principal dielectric axes, the permitivity-permeability  tensor
 of a crystal, which is always a real, symmetric matrix,  can be diagonalized as $\bK_{0}=\hbox{diag}[\epsilon_{x}, \epsilon_{y}, \epsilon_{z}, 1,1,1]$. One  type of anisotropic crystals  are  the uniaxial crystals for which $\epsilon_{x}=\epsilon_{y}=\epsilon_{\perp}\neq \epsilon_{z}=\epsilon_{\parallel}$ (if the distinguished
 direction, the optic axis,  is taken as the $z$-axis). 
  There exist two distinct
 dispersion relations for the forward modes
 \beqn
\Om^{o}&=\frac{|\bp|}{\sqrt{\epsilon_{\perp}}},\quad
 \Omega^{e}&=\sqrt{\frac{p_{3}^{2}}{\epsilon_{\perp}}+
 \frac{p_{1}^{2}+p_{2}^{2}}{\epsilon_{\parallel}}}.
 \eeqn
 The backward modes correspond to $-\Om^{e}, -\Om^{o}$. 
The corresponding  wavevector surface consists of 
 % eigenvalues of the matrix $\bK^{-1}_{0}  p_j\mbR_j$
 a sphere  and an ovaloid, a surface
 of revolution. The former 
 corresponds to an {\em ordinary} wave with a velocity independent
 of the wavevector, the latter an
 {\em extraordinary} wave with a velocity depending on the angle
 between the wavevector and the optic axis \cite{BW}. 

Let  $\bd^{o}, \bd^{e}$ be the associated 
 eigenvectors. 
Set
 $\bK_{0}^{\epsilon}=\hbox{diag}[\epsilon_{\perp},
 \epsilon_{\perp},\epsilon_{\parallel}]$ and let
 ${\mb a}^{\sigma}$ solve the following symmetric eigenvalue problem:
\beq
\label{eig2}
-\bp\times \big(\bK_{0}^{\epsilon}\big)^{-1}\bp\times
{\mb a}^{\sigma}=\big(\Om^{\sigma}\big)^{2}{\mb a}^{\sigma},\quad \sigma=e, o.
\eeq
Then the  eigenvectors
  $\bd^{\sigma}$ can be written as
 \beq
 \label{eig1}
 \bd^{\sigma}&\sim &\lt(\begin{matrix}
  -\bp\times
 {\mb a}^{\sigma}\\
 \Om^{\sigma} {\mb a}^{\sigma}
 \end{matrix}\rt),\quad\sigma= e, o.
 \eeq
 The same formula applies to the backward modes. 
 Eq. (\ref{eig2}) has the following solutions
  \[
 {\mb a}^{e}=(-p_{2}, p_{1}, 0)^\dagger,\quad
 {\mb a}^{o}=(p_{1}, p_{2}, -\frac{p_{1}^{2}+p_{2}^{2}}{p_{3}})^\dagger
 \]
 from which we deduce that the wave is linearly polarized. 
 
\subsection{Gyrotropic media: magneto-optical effect}
\label{sec:gyro}
For an isotrpic medium \cite{LLP} in motion or 
in the presence of a static external magnetic field $\bH_{\rm ext}$ the permittivity tensor $\bK^{\epsilon}_{0}$  is no longer symmetrical; it is
generally a complex Hermitian matrix. Here we consider
the simplest such constitutive relation
\beq
\label{gyrok}
\bD=\epsilon_{0}\bE-i\bg\times\bE,\quad\bB=\bH
\eeq
where  ${\mb g}=f\bH_{\rm ext}, f\in \IR, $ is the gyration vector. Equivalently, we can write 
\[
\bE=\frac{1}{\epsilon_{0}^{2}-|\bg|^{2}}
\lt(\epsilon_{0}\bD+i\bg \times \bD-\frac{1}{\epsilon_{0}}\bg\bg^{\dagger}\bD\rt). 
\]
In this case there are two distinct  forward dispersion
relations \cite{LLP}
\[
{\Om^{1}}=c_{0}\big|\bp+\frac{\Om^{1}}{2}\bg\big|,
\quad {\Om^{2}}=c_{0}\big|\bp-\frac{\Om^{2}}{2}\bg\big|
\]
 where $c_{0}=1/\sqrt{\epsilon_{0}}$.  
Clearly the wave-vector surface consists of two spheres of the same radius  but different centers. This should be contrasted with
the case of chiral media for which the wave-vector surface consists of two concentric spheres of different radii. 

The associated eigenvectors $\bd^{\sigma}, \sigma=1,2$ can be written as in (\ref{eig1})
with ${\mb a}^{\sigma}$ solving 
(\ref{eig2}) and with
 \beqn
\label{kmat}
\bK^{\epsilon}_{0}=\lt[\begin{matrix}
\epsilon_{0}&ig_{3}&-ig_{2}\\
-ig_{3}&\epsilon_{0}&ig_{1}\\
ig_{2}&-ig_{1}&\epsilon_{0}
\end{matrix}
\rt]. 
\eeqn

Let $\bg=g_{1}\hat\bp^{1}_{\perp}+g_{2}\hat\bp^{2}_{\perp}
+g_{3}\hat\bp$. We can write the three-dimensional vector ${\mb a}^{\sigma}$ as ${\mb a}^{\sigma}=\hat\bp^{1}_{\perp}+\gamma_{\sigma}\hat\bp^{2}_{\perp}$ with 
\beqn
\gamma_{\sigma}=\frac{g_{2}^{2}-g_{1}^{2}
-(-1)^{\sigma}\sqrt{(g_{1}^{2}+g_{2}^{2})^{2}+4\epsilon_{0}^{2}g_{3}^{2}}}{2(g_{1}g_{2}-i\epsilon_{0}g_{3})},\quad \sigma=1,2.
\eeqn
We see that
the wave is in general elliptically  polarized or  linearly polarized  when $\bg$ is orthogonal to the wavevector $\bp$ and circularly polarized when
$\bg$ is parallel to $\bp$.  
Again,  the simplicity of the eigenvalues implies
that depolarization is absent in
 the gyrotropic media. 

\section{Conclusion}\label{con}
The main contribution of this work is the derivation
of the 2f-RT equations (\ref{2frt}), (\ref{2frt2})
for the 2f-WD in the weak-disorder regime based
on the first Wigner-Moyal equation (\ref{wig-mo}).
All the terms in the equations can be explicitly 
calculated from the materials properties. 

Let us turn to the second Wigner-Moyal equation
(\ref{wig-mo2}) and briefly discuss its implications.
By the same multi-scale expansion, the leading order
term from eq. (\ref{wig-mo2}) is
\beqn
2\bar\om\bar\bW =i p_j\mbR_j\bL\bar\bW+{i}\bar\bW\bL  p_j\mbR_j
\eeqn
which, along with (\ref{w0}), then implies that
the wavenumber $\bp$ should be restricted to
the surface:
\[
\bar\om=\Om^{\sigma}(\bp).
\]
Denote the area element of the surface by $d \Om$.
Hence the mutual coherence in this regime is given approximately as
\beqn
\lefteqn{\lan \bu(\bx_{1},t_{1})\bu^{\dagger}(\bx_{2},t_{2})\ran}\\
&\sim&\sum_{\sigma,\alpha\zeta}\int\int e^{-i\om' t} e^{-i\tau \bar\om/\ell} \int_{\bar\om=\Om^{\sigma}(\bp)}e^{i\bp^{\dagger}(\bx_{1}-\bx_{2})/\ell}
\bar W^{\sigma}_{\alpha\zeta}(\frac{\bx_{1}+\bx_{2}}{2},\bp;\bar\om,\om')\bD^{\sigma,\alpha\zeta}(\bp,\bp) d\Om(\bp) d\bar\om d\om'
\eeqn
where the coherence matrix $\bar\bW^{\sigma}=[\bar W^{\sigma}_{\alpha\zeta}]$ satisfies
the 2f-RT equations (\ref{2frt}) and
$d\Om(\bp)$ is the area element of the surface $\bar\om=\Om^{\sigma}(\bp)$.

\begin{appendix}
\section{Derivation of Wigner-Moyal equation}
From the Maxwell equations, we have
\beq
\nn
i\om_{1}\bW&=&\frac{1}{(2\pi)^{3}}
\int e^{-i\bp^{\dagger}\by}\mbR_j\pxj\lt(\bL(\bx+\frac{\ell\by}{2})\bU_{1}\rt)
\bU^{\dagger}_{2}d\by\\
&=&\frac{2i}{\ell (2\pi)^{3}}  p_j\mbR_j\int e^{-i\bp^{\dagger}\by}
 \bL(\bx+\frac{\ell\by}{2}) \bU_{1}\bU^{\dagger}_{2}d\by\nn\\
 &&+
\frac{1}{(2\pi)^{3}} 
\int e^{-i\bp^{\dagger}\by} \mbR_{j}\bL(\bx+\frac{\ell\by}{2})
\bU_{1}\pxj\bU_{2}^{\dagger}d\by\nn
\eeq
after changing variable and integrating by parts.
Using the identity
\beq
\mbR_{j}\pxj\int e^{-i\bp^{\dagger}\by} \bL(\bx+\frac{\ell\by}{2})
\bU_{1}\bU_{2}^{\dagger}d\by
&=&
\int e^{-i\bp^{\dagger}\by} \mbR_{j}\pxj\lt[\bL(\bx+\frac{\ell\by}{2})
\bU_{1}\rt]\bU_{2}^{\dagger}d\by\nn\\
&&+ 
\int e^{-i\bp^{\dagger}\by} \mbR_{j}\bL(\bx+\frac{\ell\by}{2})
\bU_{1}\pxj\bU_{2}^{\dagger}d\by\nn
\eeq
we then obtain
\beq
\nn
i\frac{\om_{1}}{\ell}\bW&=&\frac{1}{(2\pi)^{3}}
\int e^{-i\bp^{\dagger}\by}\mbR_j\pxj\lt(\bL(\bx+\frac{\ell\by}{2})\bU_{1}\rt)
\bU^{\dagger}_{2}d\by\\
&=&\frac{i}{\ell (2\pi)^{3}}  p_j\mbR_j\int e^{-i\bp^{\dagger}\by}
 \bL(\bx+\frac{\ell\by}{2}) \bU_{1}\bU^{\dagger}_{2}d\by\nn\\
 &&+
\frac{1}{2(2\pi)^{3}} \mbR_{j}\pxj
\int e^{-i\bp^{\dagger}\by} \bL(\bx+\frac{\ell\by}{2})
\bU_{1}\bU_{2}^{\dagger}d\by. \label{a.1}
\eeq
Similarly, 
\beq
\nn
-i\frac{\om_{2}}{\ell}\bW&=&\frac{1}{(2\pi)^{3}}
\int e^{-i\bp^{\dagger}\by}\bU_{1}\nabla_{\bx}\lt(\bU^{\dagger}_{2}\bL(\bx-\frac{\ell\by}{2})\rt) \mbR_j d\by\\
&=&-\frac{i}{\ell (2\pi)^{3}}\int e^{-i\bp^{\dagger}\by}
 \bU_{1}\bU^{\dagger}_{2}\bL(\bx-\frac{\ell\by}{2})d\by   p_j\mbR_j\nn\\
 && +
\frac{1}{2(2\pi)^{3}} \nabla_{\bx}
\int e^{-i\bp^{\dagger}\by}\bU_{1}\bU_{2}^{\dagger} \bL(\bx-\frac{\ell\by}{2})
d\by\mbR_j\label{a.2}
\eeq
By the spectral representation of $\bL$ 
we write 
\beq
\label{a.3}
\frac{1}{(2\pi)^{3}}\int e^{-i\bp^{\dagger}\by}
\bL(\bx+\frac{\ell\by}{2})\bU_{1}\bU_{2}^{\dagger}d\by
&=&\int e^{i\bq^{\dagger}\bx} \widehat\bL(\bq) \bW(\bp-\frac{\ell\bq}{2})d\bq\\
\frac{1}{(2\pi)^{3}}\int e^{-i\bp^{\dagger}\by}
\bU_{1}\bU_{2}^{\dagger}\bL(\bx-\frac{\ell\by}{2})d\by
&=&\int\bW(\bp+\frac{\ell\bq}{2}) \widehat\bL(\bq) e^{i\bq^{\dagger}\bx} d\bq.\label{a.4}
\eeq
Adding or subtracting (\ref{a.1}) and (\ref{a.2}) with
(\ref{a.3})-(\ref{a.4})
we obtain the Wigner-Moyal equations.

\section{Derivation of geometrical optics equation}
%\subsection{$\cP_{2}\bar\bW$}
Consider the following term from $\cP_{2}\bar \bW$:
\beq
&&\hbox{Tr} \lt[\bE^{\tau,\xi\nu\dagger}
\mbR_{j} \pxj\lt[\bL\bar W^{\sigma}_{\alpha\zeta}
\bD^{\sigma,\alpha\zeta}\rt]\rt]
 \label{71} \\
 &=&
\hbox{Tr} \lt[\bE^{\tau,\xi\nu\dagger}
\mbR_{ j}\bL \partial_{x_{j}}\lt[\bar W^{\sigma}_{\alpha\zeta}
\bD^{\sigma,\alpha\zeta}\rt]\rt]
+\hbox{Tr} \lt[\bE^{\tau,\xi\nu\dagger}
\mbR_j\pxj\bL\bar W^{\sigma}_{\alpha\zeta}
\bD^{\sigma,\alpha\zeta}\rt]\nn\\
&=&\hbox{Tr} \lt[\bE^{\tau,\xi\nu\dagger}
\mbR_{ j}\bL \partial_{x_{j}}\bar W^{\sigma}_{\alpha\zeta}
\bD^{\sigma,\alpha\zeta}\rt]
+\hbox{Tr} \lt[\bE^{\tau,\xi\nu\dagger}
\mbR_{ j}\bL \bar W^{\sigma}_{\alpha\zeta}
\partial_{x_{j}}\bD^{\sigma,\alpha\zeta}\rt]
+\hbox{Tr} \lt[\bE^{\tau,\xi\nu\dagger}
\mbR_j\pxj\bL\bar W^{\sigma}_{\alpha\zeta}
\bD^{\sigma,\alpha\zeta}\rt]\nn
\eeq
The first term  on the right hand side of (\ref{71})   can be calculated as
\beq
\label{72}
&&\hbox{Tr} \lt[\bE^{\tau,\xi\nu\dagger}
\mbR_{ j}\bL\cdot \partial_{x_{j}}\bar W^{\sigma}_{\alpha \zeta}
\bD^{\sigma,\alpha\zeta}\rt]\\
&=&\hbox{Tr} \lt[\bE^{\tau,\xi\nu\dagger}
\partial_{p_{j}}\lt[p_{l}\mbR_{l}\bL\rt]\partial_{x_{j}}\bar W^{\sigma}_{\alpha\zeta}
\bD^{\sigma,\alpha\zeta}\rt]\nn\\
&=&\hbox{Tr} \lt[\bb^{\tau,\nu}\partial_{p_{j}}\lt[\bb^{\tau,\xi\dagger}
p_{l}\mbR_{l}\bL\rt]\partial_{x_{j}}\bar W^{\sigma}_{\alpha\zeta}
\bD^{\sigma,\alpha\zeta}\rt]-
\hbox{Tr} \lt[\bb^{\tau,\nu}\partial_{p_{j}}\lt[\bb^{\tau,\xi\dagger}
\rt]
p_{l}\mbR_{l}\bL\partial_{x_{j}}\bar W^{\sigma}_{\alpha \zeta}
\bD^{\sigma,\alpha\zeta}\rt]\nn\\
&=&\hbox{Tr} \lt[\bE^{\tau,\xi\nu\dagger}\partial_{p_{j}}\Om^{\tau}\partial_{x_{j}}\bar W^{\sigma}_{\alpha\zeta}
\bD^{\sigma,\alpha\zeta}\rt]+
\hbox{Tr} \lt[\bb^{\tau,\nu}\partial_{p_{j}}\lt[\bb^{\tau,\xi\dagger}
\rt]
\lt[\Om^{\tau}-p_{l}\mbR_{l}\bL\rt]\partial_{x_{j}}\bar W^{\sigma}_{\alpha\zeta}
\bD^{\sigma,\alpha\zeta}\rt]\nn\\
&=&\delta_{\tau\delta}\delta_{\xi\alpha}
\delta_{\nu\zeta}\nabla_{\bp}\Om^{\tau}\cdot\nabla_{\bx}\bar W^{\tau}_{\xi\nu}\nn
\eeq
using the eigenvector property and (\ref{lr})
while the last term on the right hand side of (\ref{71}) is
\beq
\hbox{Tr} \lt[\bE^{\tau,\xi\nu\dagger}
\mbR_j\pxj\bL\bar W^{\sigma}_{\alpha\zeta}
\bD^{\sigma,\alpha\zeta}\rt]\label{73}
%&=&\bb^{\tau,\xi\dagger}\mbR_j\pxj\bL
%\bar W^{\tau,\alpha\nu}\bd^{\tau,\alpha}\nn\\
&=&\delta_{\tau\sigma}\delta_{\nu\zeta}\bb^{\tau,\xi\dagger}
\mbR_{j}\pxj \bL\bd^{\tau,\alpha}
\bar W^{\tau}_{\alpha\nu}.
\commentout{
&=&\delta_{\tau\sigma}\delta_{\nu\zeta}\lt[\nabla_{\bp}\cdot\nabla_{\bx}\lt[\bb^{\tau,\xi\dagger}
  p_j\mbR_j\bL\rt]
\bar W^{\tau,\alpha\nu}\bd^{\tau,\alpha}
-\nabla_{\bp}\cdot\nabla_{\bx}\bb^{\tau,\xi\dagger}
  p_j\mbR_j\bL
\bar W^{\tau,\alpha\nu}\bd^{\tau,\alpha}\rt]\nn\\
&=&\delta_{\tau\sigma}\delta_{\nu\zeta}\lt[\nabla_{\bp}\cdot\nabla_{\bx}\lt[\Om^{\tau}\bb^{\tau,\xi\dagger}\rt]
\bar W^{\tau,\alpha\nu}\bd^{\tau,\alpha}
-\nabla_{\bp}\cdot\nabla_{\bx}\bb^{\tau,\xi\dagger}
  p_j\mbR_j\bL
\bar W^{\tau,\alpha\nu}\bd^{\tau,\alpha}\rt]\nn\\
&=&\delta_{\tau\sigma}\delta_{\nu\zeta}\delta_{\xi\alpha}\nabla_{\bp}\cdot\nabla_{\bx}\Om^{\tau}
\bar W^{\tau,\xi\nu}.\nn
}
\eeq
We turn to the middle term on the right hand side of (\ref{71}). We have the following calculation.
\beq
\label{74}
&&\sum_{\sigma,\alpha,\zeta}\hbox{Tr} \lt[\bE^{\tau,\xi\nu\dagger}
\mbR_{j}\bL \bar W^{\sigma}_{\alpha\zeta}
\lt[\partial_{x_{j}}\bd^{\sigma,\alpha}\bd^{\sigma,\zeta\dagger}
+\bd^{\sigma,\alpha}\partial_{x_{j}}\bd^{\sigma,\zeta\dagger}\rt]\rt]\\
&=&
\bb^{\tau,\xi\dagger}\mbR_{j}\bL\partial_{x_{j}}\bd^{\tau,\alpha}
\bar W^{\tau}_{\alpha\nu}+\sum_{\sigma,\alpha,\zeta}\hbox{Tr}\lt[\bb^{\tau,\nu}\partial_{p_l}
\lt[\bb^{\tau,\xi\dagger}  p_j\mbR_j\bL\rt]
\bar W^{\sigma}_{\alpha\zeta}\bd^{\sigma,\alpha}
\partxl \bd^{\sigma,\zeta\dagger}\rt]\nn\\
&&-\sum_{\sigma,\alpha,\zeta}\hbox{Tr}\lt[\bb^{\tau,\nu}\partpl
\bb^{\tau,\xi\dagger} \bar W^{\sigma}_{\alpha\zeta} p_j\mbR_j\bL
\bd^{\sigma,\alpha}
\partxl \bd^{\sigma,\zeta\dagger}\rt]\nn\\
&=&
\bb^{\tau,\xi\dagger}\mbR_{j}\bL\partial_{x_{j}}\bd^{\tau,\alpha}
\bar W^{\tau}_{\alpha\nu}+\sum_{\sigma,\alpha,\zeta}\hbox{Tr}\lt[\bb^{\tau,\nu}\partpl
\lt[\Om^{\tau}\bb^{\tau,\xi\dagger}\rt]
\bar W^{\sigma}_{\alpha\zeta}\bd^{\sigma,\alpha}
\partxl\bd^{\sigma,\zeta\dagger}\rt]\nn\\
&&-\sum_{\sigma,\alpha,\zeta}\hbox{Tr}\lt[\bb^{\tau,\nu}\partpl\bb^{\tau,\xi\dagger}
\bar W^{\sigma}_{\alpha\zeta}\Om^{\sigma}\bd^{\sigma,\alpha}
\partxl \bd^{\sigma,\zeta\dagger}\rt]\nn\\
&=&\sum_{\alpha}\bb^{\tau,\xi\dagger}\mbR_{j}\bL\partial_{x_{j}}\bd^{\tau,\alpha}\bar
W^{\tau}_{\alpha\nu}
+\sum_{\zeta}\partial_{p_{j}}
\Om^{\tau}
\partial_{x_{j}}\bd^{\tau,\zeta\dagger}\bb^{\tau,\nu}\bar W^{\tau}_{\xi\zeta}\nn\\
&&+\sum_{\sigma,\alpha,\zeta}\lt(\Om^{\tau}-\Om^{\sigma}
\rt) \bar W^{\sigma}_{\alpha\zeta}\partial_{p_{j}}\bb^{\tau,\xi\dagger}\bd^{\sigma,\alpha}\partial_{x_{j}}\bd^{\sigma,\zeta\dagger}\bb^{\tau,\nu}.\nn
\eeq
The second part in $\cP_{2}\bar \bW$ can be calculated in
  the same manner. The counterpart of (\ref{72})
  yields exactly the same expression as the
  right hand side of (\ref{72})
 while 
 the counterparts of  (\ref{73}) and (\ref{74}) yield, respectively
 \beqn
\sum_{\zeta} \bar W^{\tau}_{\xi\zeta}\bd^{\tau,\zeta\dagger}\pxj \bL\mbR_{j}
 \bb^{\tau,\nu}
 \eeqn
 and 
 \beq
\label{75}
&&\sum_{\sigma,\alpha,\zeta}\hbox{Tr} \lt[\bE^{\tau,\xi\nu\dagger}
 \bar W^{\sigma}_{\alpha\zeta}
\lt[\partial_{x_{j}}\bd^{\sigma,\alpha}\bd^{\sigma,\zeta}
+\bd^{\sigma,\alpha}\partial_{x_{j}}\bd^{\sigma,\zeta}\rt]\bL\mbR_{j}\rt]\\
&=&\sum_{\zeta}\bar W^{\tau}_{\xi\zeta}
\partial_{x_{j}}\bd^{\tau,\zeta\dagger}\bL\mbR_{j}\bb^{\tau,\nu}
+\sum_{\alpha}
\bb^{\tau,\xi\dagger}\partxl\bd^{\tau,\alpha}\partpl\Om^{\tau}\bar W^{\tau}_{\alpha\nu} 
\nn\\
&&+\sum_{\sigma,\alpha,\zeta}\lt(\Om^{\tau}-\Om^{\sigma}
\rt) \bar W^{\sigma}_{\alpha\zeta}\bb^{\tau,\xi\dagger}\partxl\bd^{\sigma,\alpha}\bd^{\sigma,\zeta\dagger}\partpl\bb^{\tau,\nu}.\nn
\eeq

Next let us turn to the first part in $\cP_{1}\bar\bW$:
\beq
\label{76}
&&\sum_{\sigma,\alpha,\zeta}\hbox{Tr} \lt[\bE^{\tau,\xi\nu\dagger}  p_j\mbR_j\nabla_{\bx}
\bL\cdot \nabla_{\bp}\lt[\bar W^{\sigma}_{\alpha\zeta}
\bd^{\sigma,\alpha}\bd^{\sigma,\zeta\dagger}\rt]\rt]\\
&=&\sum_{\alpha}\bb^{\tau,\xi\dagger}  p_j\mbR_j\nabla_{\bx}
\bL\cdot\nabla_{\bp}\bar W^{\tau}_{\alpha\nu}
\bd^{\tau,\alpha}
+\sum_{\sigma,\alpha,\zeta}\hbox{Tr} \lt[\bE^{\tau,\xi\nu\dagger}  p_j\mbR_j\partial_{x_{j}}
\bL\bar W^{\sigma}_{\alpha\zeta}\partial_{p_{j}}\lt[
\bd^{\sigma,\alpha}\bd^{\sigma,\zeta\dagger}\rt]\rt].\nn
\eeq
The first term on the right hand side of  (\ref{76}) equals 
\beq
\label{77}
\sum_{\alpha}\partial_{x_{j}}\lt(\Om^{\tau}\bb^{\tau,\xi\dagger}\rt)
\bd^{\tau,\alpha}\partial_{p_{j}}\bar W^{\tau}_{\alpha\nu}
-\sum_{\alpha}\partial_{x_{j}}\bb^{\tau,\xi\dagger}  p_j\mbR_j\bL
\bd^{\tau,\alpha}\partial_{p_{j}}\bar W^{\tau}_{\alpha\nu} 
&=&\nabla_{\bx}\Om^{\tau}\cdot\nabla_{\bp}
\bar W^{\tau}_{\xi\nu}\nn
\eeq
while the second term can be calculated as 
\beq
&&\sum_{\sigma,\alpha,\zeta}\partial_{x_{j}}\lt(\Om^{\tau}\bb^{\tau,\xi\dagger}\rt)\bar W^{\sigma}_{\alpha\zeta}\partial_{p_{j}}\lt[
\bd^{\sigma,\alpha}\bd^{\sigma,\zeta\dagger}\rt]\bb^{\tau,\nu}
-\sum_{\sigma,\alpha,\zeta}\partial_{x_{j}}\bb^{\tau,\xi\dagger}  p_j\mbR_j\bL\bar W^{\sigma}_{\alpha\zeta}\partial_{p_{j}}\lt[
\bd^{\sigma,\alpha}\bd^{\sigma,\zeta\dagger}\rt]\bb^{\tau,\nu}
\nn\\
&=& \sum_{\alpha}
\bb^{\tau,\xi\dagger}\partial_{p_{j}}\bd^{\tau,\alpha}\partial_{x_{j}}\Om^{\tau}\bar W^{\tau}_{\alpha\nu}
+\sum_{\zeta}\bar W^{\tau}_{\xi\zeta}
\partial_{x_{j}}\Om^{\tau} \partial_{p_{j}}
\bd^{\tau,\zeta\dagger}\bb^{\tau,\nu}+\sum_{\alpha}\Om^{\tau}\bar W^{\tau}_{\alpha\nu}\partial_{x_{j}}
\bb^{\tau,\xi\dagger}\partial_{p_{j}}\bd^{\tau,\alpha}\nn\\
&&+\sum_{\sigma,\alpha,\zeta}\lt(\Om^{\tau}-\Om^{\sigma}\rt)\bar W^{\sigma}_{\alpha\zeta}\partial_{x_{j}}\bb^{\tau,\xi\dagger}
\bd^{\sigma,\alpha}\partial_{p_{j}}\bd^{\sigma,\zeta\dagger}
\bb^{\tau,\nu}-\sum_{\alpha}
\bar W^{\tau}_{\alpha\nu}
\partial_{x_{j}}\bb^{\tau,\xi\dagger}  p_j\mbR_j\bL
\partial_{p_{j}}\bd^{\tau,\alpha}.\label{78}
\eeq
The last term on the right hand side of (\ref{78}) can be
further expressed as
\beq
\label{79}
&&-\sum_{\alpha}
\bar W^{\tau}_{\alpha\nu}
\partial_{x_{j}}\bb^{\tau,\xi\dagger}\partial_{p_{j}}\lt[  p_j\mbR_j\bL
\bd^{\tau,\alpha}\rt]
+\sum_{\alpha}
\bar W^{\tau}_{\alpha\nu}
\partial_{x_{j}}\bb^{\tau,\xi\dagger}\mbR_{j}\bL
\bd^{\tau,\alpha}\\
&=&-\sum_{\alpha}
\bar W^{\tau}_{\alpha\nu} \partial_{p_{j}}
\Om^{\tau}
\partial_{x_{j}}\bb^{\tau,\xi\dagger}\bd^{\tau,\alpha}
-\sum_{\alpha}
\bar W^{\tau}_{\alpha\nu}\Om^{\tau}
\partial_{x_{j}}\bb^{\tau,\xi\dagger} \partial_{p_{j}}
\bd^{\tau,\alpha}\nn\\
&&+\sum_{\alpha}
\partial_{x_{j}}\bb^{\tau,\xi\dagger}\mbR_{j}\bL 
\bd^{\tau,\alpha} \bar W^{\tau}_{\alpha\nu}.\nn
\eeq
For the second part in $\cP_{1}\bar \bW$, the counterpart
of (\ref{77}) yields exactly the same expression as (\ref{77})
while the counterpart of (\ref{78}) yields
\beq
&&\sum_{\alpha}\bb^{\tau,\xi\dagger}\ppj \bd^{\tau,\alpha}\pxj\Om^{\tau}
\bar W^{\tau}_{\alpha\nu}+\sum_{\zeta}
\bar W^{\tau}_{\xi\zeta}\pxj \Om^{\tau}\ppj \bd^{\tau,\zeta\dagger}\bb^{\tau,\nu}\nn\\
&&+\sum_{\sigma,\alpha,\zeta}\lt(\Om^{\tau}-\Om^{\sigma}\rt)
\bar W^{\sigma}_{\alpha\zeta}
\bb^{\tau,\xi\dagger}\ppj \bd^{\sigma,\alpha}\bd^{\sigma,\zeta\dagger}
\pxj \bb^{\tau,\nu} 
+\bar W^{\tau}_{\xi\zeta}\bd^{\tau,\zeta\dagger}\bL
\mbR_{j}\pxj \bb^{\tau,\nu}-
\bar W^{\tau}_{\xi\zeta}\ppj \Om^{\tau} \bd^{\tau,\xi\dagger}
\pxj \bb^{\tau,\nu}.\nn
\eeq
In the final expression, 
 many of the above terms in $\cP_{1}\bar \bW-\cP_{2}\bar \bW$ cancel. For instance, all the terms involving 
 $(\Om^{\tau}-\Om^{\sigma})$, 
 all the terms involving $\bar W^{\tau}_{\xi\zeta}\ppj \Om^{\tau}, \bar W^{\tau}_{\alpha\nu}\ppj \Om^{\tau}$ and all the terms involving $\Om^{\tau}
 \bar W^{\tau}_{\xi\zeta}, \Om^{\tau}\bar W^{\tau}_{\alpha\nu}$ cancel. Using
 the fact $\bb^{\tau,\alpha}=\bL\bd^{\tau,\alpha}$
 and some algebra
% and that $ \partial_{p_{j}}
%\bd^{\tau,\zeta\dagger}\bb^{\tau,\nu}=
%-
%\bd^{\tau,\zeta\dagger}\ppj \bb^{\tau,\nu}$
we obtain
 \beq
&&\hbox{Tr}\lt[\bE^{\tau,\xi\nu\dagger}\cP_{1}\bar \bW\rt]-
\hbox{Tr}\lt[\bE^{\tau,\xi\nu\dagger}\cP_{2}\bar \bW \rt]-\nabla_{\bx}\Om^{\tau}\cdot\nabla_{\bp} \bar W^{\tau}_{\xi\nu}+\nabla_{\bp}\Om^{\tau}\cdot
\nabla_{\bx}\bar W^{\tau}_{\xi\nu}\\
& =&
\sum_{\alpha}\bb^{\tau,\xi\dagger}\partial_{p_{j}}\bd^{\tau,\alpha}\partial_{x_{j}}\Om^{\tau}\bar W^{\tau}_{\alpha\nu}
+\sum_{\zeta}\bar W^{\tau}_{\xi\zeta}
\partial_{x_{j}}\Om^{\tau} 
\ppj\bd^{\tau,\zeta\dagger}\bb^{\tau,\nu}
+\frac{1}{2}\sum_{\alpha}\partial_{x_{j}}\bb^{\tau,\xi\dagger}\mbR_{j} 
\bb^{\tau,\alpha} \bar W^{\tau}_{\alpha\nu}
\nn\\
&&
-\frac{1}{2}\sum_{\alpha}\bb^{\tau,\xi\dagger}\mbR_{j}\pxj\bb^{\tau,\alpha}\bar W^{\tau}_{\alpha\nu}+\frac{1}{2}\sum_{\zeta}\bar W^{\tau}_{\xi\zeta}\bb^{\tau,\zeta\dagger}
\mbR_{j}\pxj \bb^{\tau,\nu}
-\frac{1}{2}\sum_{\zeta}\bar W^{\tau}_{\xi\zeta}\pxj \bb^{\tau,\zeta\dagger}
\mbR_{j}\bb^{\tau,\nu}. \nn
\eeq

\end{appendix}

\end{document}